\documentclass[reprint]{revtex4-1}

\usepackage[usenames,dvipsnames]{color}
\usepackage{amssymb}
\usepackage{amsmath}
\usepackage{hyperref}
\usepackage{graphicx}
\usepackage{subfigure}
\usepackage{xspace}

\usepackage{lipsum}

\newcommand{\mthr}{\mathrm}
\newcommand{\on}{\overline n}

\begin{document}

\title{Surface-triggered cascade reactions between DNA linkers direct self-assembly of colloidal crystals of controllable thickness}

\author{Pritam Kumar Jana$^{\ast}$\textit{$^{a}$} and Bortolo Matteo Mognetti$^{\ddag}$\textit{$^{a}$}}

\address{
$^{a}$Universit\'e Libre de Bruxelles (ULB), Interdisciplinary Center for Nonlinear Phenomena and Complex Systems, Campus Plaine, CP 231, Blvd.\ du Triomphe, B-1050 Brussels. \\
$^{\ast}$Pritam.Kumar.Jana@ulb.ac.be $^{\ddag}$bmognett@ulb.ac.be
}

\begin{abstract}
Functionalizing colloids with reactive DNA linkers is a versatile way of programming self-assembly. DNA selectivity provides direct control over colloid-colloid interactions allowing the engineering of structures such as complex crystals or gels. However, self-assembly of localized and finite structures remains an open problem with many potential applications. In this work, we present a system in which functionalized surfaces initiate a cascade reaction between linkers leading to self-assembly of crystals with a controllable number of layers. Specifically, we consider colloidal particles functionalized by two families of complementary DNA linkers with mobile anchoring points, as found in experiments using emulsions or lipid bilayers. In bulk, intra-particle linkages formed by pairs of complementary linkers prevent the formation of inter-particle bridges and therefore colloid-colloid aggregation. However, colloids interact strongly with the surface given that the latter can destabilize intra-particle linkages. When in direct contact with the surface, colloids are activated, meaning that they feature more unpaired DNA linkers ready to react. Activated colloids can then capture and activate other colloids from the bulk through the formation of inter-particle linkages. Using simulations and theory, validated by existing experiments, we clarify the thermodynamics of the activation and binding process and explain how particle-particle interactions, within the adsorbed phase, weaken as a function of the distance from the surface. The latter observation underlies the possibility of self-assembling finite aggregates with controllable thickness and flat solid-gas interfaces. Our design suggests a new avenue to fabricate heterogeneous and finite structures.
\end{abstract} 

\maketitle

\section{Introduction}

Many recent contributions have unveiled the advantages of using complementary single-stranded (ss) DNA oligomers tethered to colloidal particles to program self-assembly \cite{Mirkin_Nature_1996,Alivisatos_Nature_1996,Jones_Science_2015}. The selectivity of Watson-Crick base pairing underlies most of the functionalities and responsive behaviors achieved using DNA \cite{Jones_Science_2015}. For instance, DNA has been used to self-assemble colloidal crystals lacking molecular analog \cite{Macfarlane_AngChem_52} or featuring optical bandgaps \cite{Ducrot_NatMat_2017,Wang_NatComm_2017,Liu_Science_2016}, or engineer bigels \cite{Varrato_PNAS_2012} and re-entrant phase behaviors \cite{rogers-manoharan,Angioletti-Uberti_NMat_2012}.
\\
Recently, systems of particles functionalized by mobile linkers tipped by reactive sites received a lot of attention \cite{Beales_SoftMatter_2011,Pontani_PNAS_2012,Meulen_JACS_2013,chakraborty2017colloidal,Hadorn_PNAS_2012,Parolini_NatComm_2015,hu2018entropy}.  In these systems, DNA oligomers conjugated to hydrophobic tags are tethered, for instance, to lipid bilayers \cite{Meulen_JACS_2013,chakraborty2017colloidal}. 
Lipid vesicles functionalized by mobile ligands are currently employed in nanotechnological platforms to mimic biological functionalities like cell adhesion and recognition \cite{Pontani_PNAS_2012,Parolini_NatComm_2015}. In self-assembly, mobile ligands facilitate annealing of crystal defects \cite{Meulen_JACS_2013} and could allow remote control over the valency of the aggregates \cite{Feng_SoftMatter_2013,Angioletti-Uberti_PRL_2014}. 
In this respect, colloidal supported bilayers \cite{TROUTIER20071, rinaldin2018colloidal} functionalized with DNA linkers \cite{rinaldin2018colloidal} are a new generation of a particularly versatile type of rigid and monodisperse building blocks with enhanced programmability given by the mobility of the binders \cite{Angioletti-Uberti_PRL_2014}.
\\
So far DNA mediated interactions have been used almost exclusively to fabricate extended colloidal structures. However, the ability to constrain self-assembly spatially is central in many nanotechnological applications including encapsulation and development of point-of-care devices \cite{Chapman2015}. Localized self-assembly is also important in biology where, for instance, many cell functionalities rely on dynamic compartmentalized environments (e.g.~Refs.~\cite{Conduit_NatRevMCB_2015,Case_NatCellBio_2015}). Developing bottom-up methods for controlled surface coating is also a key problem in chemistry \cite{Vigier_AngChem_57}.
\begin{figure}[ht!]
\vspace{0.5cm}
\centering
\includegraphics[width=8.46cm]{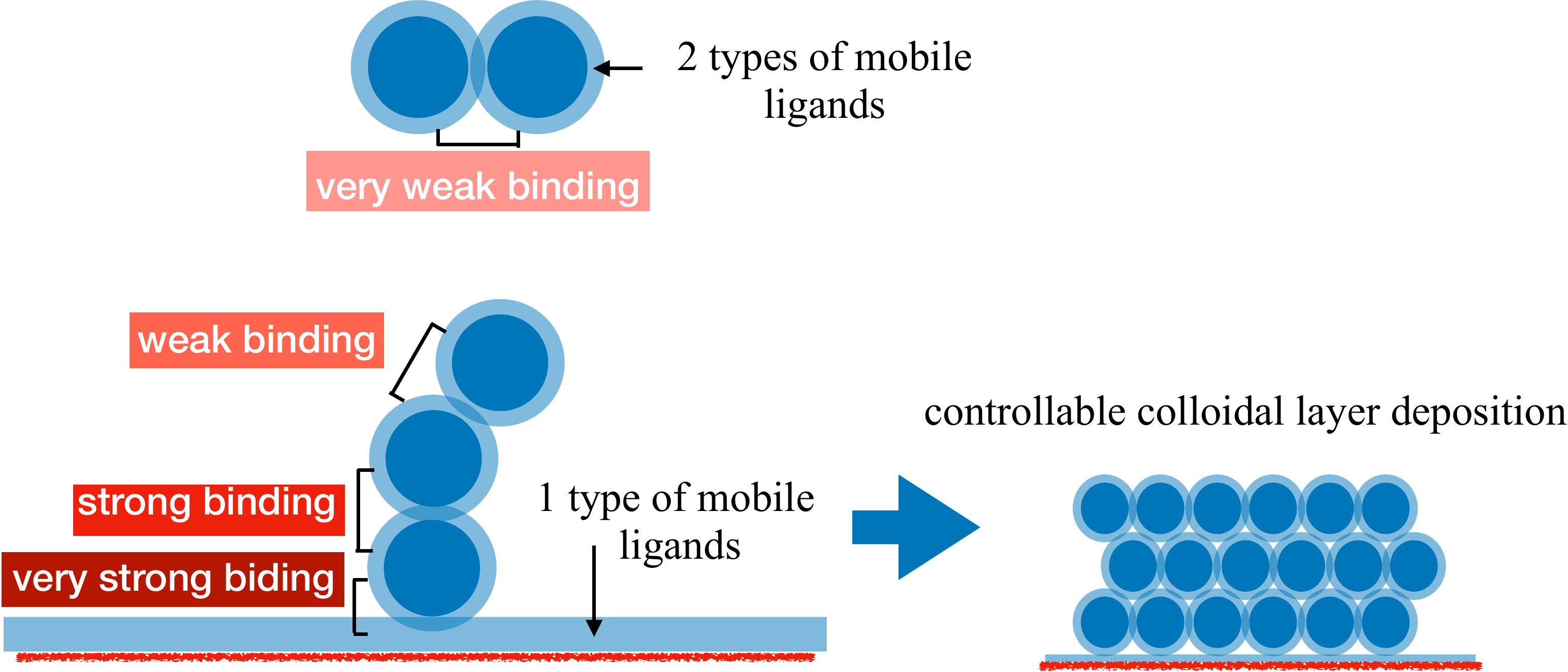} 
%\vspace{-0.5cm}
\caption{Self--assembly of finite aggregates directed by functionalized surfaces. (left) Particles carrying two types of DNA linkers strongly bind to surfaces functionalized by a single type of receptor. In bulk, particle-particle interactions are weak resulting in a stable gas phase. In our design, particle-particle interactions are magnified at the surface and sharply decrease with the particle-surface distance (left). Such a hierarchy of interactions leads to self-assembly of crystals comprising a desired number of layers (right).\label{Fig:Intro}}
\end{figure}
\\
In this contribution, we show how to use DNA to yield localized self--assembly of colloidal structures. In particular, we study a biomimetic system comprising colloidal supported bilayers functionalized by two types of complementary mobile linkers interacting with surfaces carrying a single type of DNA linker. In most systems, reactions between oligomers tethered to different particles are, to a good extent, independent events. Instead, in the present work, the surface initiates a cascade reaction between DNA linkers that propagate concomitantly with colloidal aggregation. In bulk, DNA linkers predominantly form intra--particle loops while the probability of forming inter--particle bridges remains negligible resulting in weak particle-particle interactions and stable gas phases, Fig.~\ref{Fig:Intro}. Instead, the unpaired linkers tethered to the surface can easily stabilize colloidal particles through the formation of particle-surface bridges. Once bound to the surface, particles become activated and display a higher number of free DNA linkers appearing as a side product of the reaction leading to the formation of particle--surface bridges. Importantly, activated colloids can attract and activate others colloids from the bulk. This process triggers a domino effect leading to the self--assembly of colloidal crystals at the surface, Fig.~\ref{Fig:Intro}. We use theoretical modeling, supported by Brownian dynamics simulations, to explain how enhanced particle--particle interactions at the surface is an entropic effect mainly controlled by the number of linkers {\em per} particle and the relative statistical weight of forming inter--particle and intra--particle linkages. Importantly, the domino effect does not propagate indefinitely. Instead, colloid-colloid interactions sharply decrease with the distance between newly activated colloids and the surface, Fig.~\ref{Fig:Intro}. This observation explains the flatness of the solid-fluid interfaces, a missing result in wetting phenomena where roughness and thickness are correlated quantities \cite{bonn2009wetting}. 
As compared to existing protocols leading to colloidal layer deposition (e.g.~Refs.~\cite{trau1996field,reculusa2003synthesis}), our method provides direct control over the number of deposited layers and does not require external intervention during aggregation. Such property arises from the possibility of controlling particle-particle interactions at different particle-surface distances (Fig.~\ref{Fig:Intro}) using design parameters. 
\\
Our design is robust as proven by state-of-the-art simulations of self-assembly directed by ligand-receptor complexation.  We prove the reliability of our methods using recent experiments that investigated the stability of suspensions of colloids featuring competition between bridges and loops in bulk \cite{Bachmann_SoftMatt_2016}.  Beyond addressing an important technological problem, this paper suggests new routes leading to the fabrication of localized structures and new responsive behaviors at functionalized surfaces.

\begin{figure}[h]
\vspace{0.5cm}
\centering
\includegraphics[width=8.5cm]{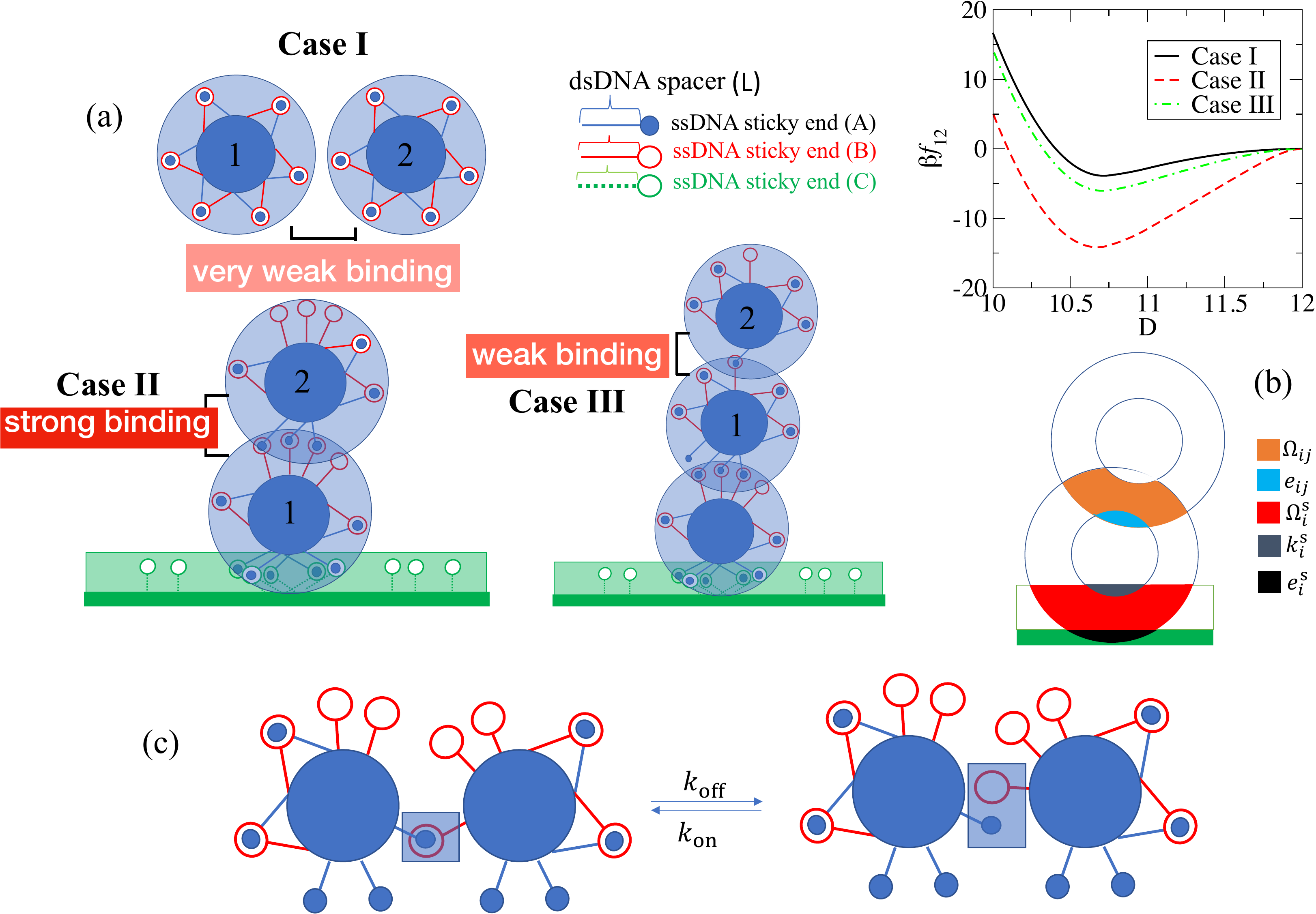} 
%\vspace{-0.5cm}
\caption{\small Functionalised surfaces re-program particle-particle interactions. $(a)$ Intra--particle loops prevent particle aggregation in bulk (Case I). Receptors displayed by the surface activate the colloids by destabilizing intra--particle loops. Activated colloids can bind and activate other colloids from the bulk through the formation of inter--particle bridges (Case II and III). The plot reports particle--particle effective interactions ($f_{12}$) at the surface and in bulk as a function of the distance $D$ between particle 1 and particle 2 (tagged as in Case I-III). In this figure, $N_\mthr{L}=50$, $\beta \Delta G_0= -9$, $\beta \Delta G^\mthr{s}_0=-10$, while the receptor density is equal to $\sigma_R=1.8\,L^{-2}$. $(b)$ Definition of the configurational volumes entering the calculation of the entropic terms regulating the number of linkages. $(c)$ Linkages form/break according to $on/off$ rates calculated using biochemical data and configurational terms (see panel $b$).\label{Fig:System} }
\end{figure}

\section{Theoretical and simulation methods}
Figs.~\ref{Fig:System}$a$ and \ref{Fig:System}$b$ present the design that we use to reproduce the peculiar collective behaviors anticipated by Fig.~\ref{Fig:Intro}. We consider rigid colloidal particles of radius $R$ functionalized by two types of linkers (or ligands), A and B, freely moving on the surfaces of the particles. Each particle carries $N_\mthr{L}$ ligands of each type, while the surface is decorated with $N_\mthr{C}$ mobile linkers (or receptors). In our model, A can bind to B and C, while C and B do not pair. Therefore, the possible linkages featured by the system are inter-particle loops, particle-particle bridges, and particle-surface bridges (see Fig.~\ref{Fig:System}$a$). Van der Meulen and Leunissen first studied silica particles coated with DNA linkers with a hydrophobized head immersed in the bilayers covering the colloids \cite{Meulen_JACS_2013}. The group of Kraft is currently studying the potentialities of colloidal supported lipid bilayers functionalized by DNA for self-assembly \cite{rinaldin2018colloidal,C6NR08069C}. In our design, both ligands and receptors comprise short rods of double-stranded (ds) DNA of length $L$ tipped by reactive single-stranded (ss) DNA sequences (see legend of Fig.~\ref{Fig:System}). In this study, we use $R=5\cdot L$ and $L$ is chosen as the unit length.
\\
As it is often the case in soft-matter, interactions between complex building blocks include entropic contributions due to ensemble averages over all possible configurations of the system compatible with a given particle-particle (or particle-surface) distance. For multivalent interactions, entropic terms are due to the configurational constraints that surface anchored binders need to satisfy (coalescence of the endpoints and, for mobile constructs, colocalization of the tethering points), and combinatorial factors counting the possible ways of making a certain number of linkages \cite{Dreyfus_PRL_2009,Martinez-Veracoechea_PNAS_2011,Angioletti-Uberti_PCCP_2016}. 
In Sec.~\ref{Sec:multi} we derive the multivalent free energy $F$ controlling particle--particle and particle--surface interactions \cite{Angioletti-Uberti_PRL_2014,Petitzon_SoftMatt_2016}, while in Sec.~\ref{Sec:Sim} we detail the simulation methods.

\subsection{Derivation of the multivalent free energy}\label{Sec:multi}
The partition function of $N_\mathrm{p}$ colloids carrying $N_\mathrm{L}$ A and B ligands facing a surface decorated with $N_\mthr{C}$ receptors of type C (see Fig.~\ref{Fig:System}) reads as 
\begin{eqnarray}
Z&=&\frac{1}{ N_\mathrm{p}!} \int d\{{\bf{r}}\} \sum_{\{ n\}} {\cal Z} (\{n\},\{{\bf{r}}\}) 
\nonumber \\
&=&\frac{1}{ N_\mathrm{p}!}\int d\{{\bf{r}}\} \sum_{\{ n\}} e^{-\beta {\cal F}_\mathrm{multi} (\{n\},\{{\bf{r}}\})}
\label{Equ:Z0}
\end{eqnarray}
where $\{\bf{r}\}$ denotes the cartesian coordinates of the particles ($\{\bf{r}\}=\{ {\bf r}_1 \cdots {\bf r}_{N_p} \}$) and $\{ n\}$ ($\{n\}=\{n^\mathrm{AB}_{ii},\, n^\mathrm{AB}_{ij},\, n^\mathrm{BA}_{ij},\, n^\mathrm{AC}_i\}$ with $i,j=1,\cdots N_\mathrm{p}$) the ensemble of intra-particle ($n^\mathrm{AB}_{ii}$), inter-particle ($n^\mathrm{AB}_{ij}=n^\mathrm{BA}_{ji}$ and $i\neq j$), and particle-surface ($n^\mathrm{AC}_i$) linkages. $n^\mathrm{AB}_{ij}$/$n^\mathrm{BA}_{ij}$ is the number of bridges between particle $i$ and $j$ resulting from binding linkers of type $A$/$B$ tethered to particle $i$ with linkers of type $B$/$A$ tethered to $j$. ${\cal Z}$ and ${\cal F}_\mathrm{multi}$ are, respectively, the partition function of the system and the multivalent free energy at given $\{ {\bf r} \}$ and $\{ n\}$. ${\cal Z}$ comprises combinatorial terms, counting the number of ways of making $\{ n \}$ linkages, and terms linked to the hybridization free energy of DNA reactive sequences \cite{SantaLucia_PNAS_1998}.
In Sec.~S1 of the SI we adapt the calculations of Refs. \cite{Angioletti-Uberti_PRL_2014,Petitzon_SoftMatt_2016,Martinez-Veracoechea_PNAS_2011} to the system of Fig.~\ref{Fig:System} and report the explicit expressions of ${\cal Z}$ and ${\cal F}_\mthr{multi}$.
\\
At given colloid positions, $\{ {\bf r} \}$, the most likely number of linkages featured by the system, $\{ \overline n \}$, are calculated by maximizing the multivalent free energy 
\begin{eqnarray}
\frac{\partial }{\partial \{ n\} } {\cal F}_\mathrm{multi} (\{n\})|_{\{n\} = \{ \on \}} = 0 \, .
\label{Equ:Saddle}
\end{eqnarray}
Eq.~\ref{Equ:Saddle}, along with the expression of ${\cal F}_\mathrm{multi}$ (see SI Sec.~1), lead to the chemical equilibrium equations for the most likely number of linkages at given $\{ {\bf r} \}$
\begin{eqnarray}
\on^{\mathrm{AB}}_{ii} &=& \on^\mathrm{A}_i \on^\mathrm{B}_i \exp[-\beta \Delta G_{ii} (\{ {\bf{r}}\})]
\nonumber\\
\on^{\mathrm{AB}}_{ij} &=& \on^\mathrm{A}_i \on^\mathrm{B}_j \exp[-\beta \Delta G_{ij} (\{ {\bf{r}}\})]
\nonumber\\
\on^{\mathrm{BA}}_{ij} &=& \on^\mathrm{B}_i \on^\mathrm{A}_j \exp[-\beta \Delta G_{ij} (\{ {\bf{r}}\})]
\nonumber\\
\on^{\mathrm{AC}}_{i}  &=& \on^\mathrm{A}_i \on^\mathrm{C}_\mathrm{s} \exp[-\beta \Delta G^\mathrm{s}_{i} (\{ {\bf{r}}\})] \, \, ,
\label{Equ:linkages}
\end{eqnarray}
where $\on_i^\mathrm{A}$, $\on_i^\mathrm{B}$, and $\on^\mathrm{C}_\mathrm{s}$ denote the number of free (unbound) linkers. In Eqs.~\ref{Equ:linkages} we defined the hybridization free energies of making inter--particle, intra--particle, and particle--surface linkages with $\Delta G_{ii}$, $\Delta G_{ij}$, and $\Delta G^\mathrm{s}_i$, respectively. The hybridization free energies comprise the binding free energy of the reactive oligomers free in solutions \cite{SantaLucia_PNAS_1998} ($\Delta G_0$ and $\Delta G_0^\mathrm{s}$ for the dimerization of $A$ with $B$ and $A$ with $C$, respectively) augmented by configurational contributions as follows   \cite{Petitzon_SoftMatt_2016,Angioletti-Uberti_PRL_2014}
\begin{eqnarray}
\beta \Delta G_{ii} (\{\bf{r}\}) &=& \beta \Delta G_0 -  \log \frac{1 }{ \rho_0 \Omega_i(\{ {\bf{r}}\})}
\label{Equ:DG}
\\
\beta \Delta G_{ij} (\{\bf{r}\}) &=& \beta \Delta G_0 -  \log \frac{\Omega_{ij}(\{ {\bf{r}}\}) }{ \rho_0{ \Omega_i (\{ {\bf{r}}\})\Omega_j(\{ {\bf{r}}\})}}
\nonumber \\
\beta \Delta G_{i}^\mathrm{s} (\{\bf{r}\}) &=& \beta \Delta G_0^{\mathrm{s}} -  \log \frac{\Omega_{i}^{\mathrm{s}}(\{ {\bf{r}}\}) }{ \rho_0 { \Omega_i (\{ {\bf{r}}\}) \Omega^{\mathrm{s}}}(\{ {\bf{r}}\})} \, ,
\nonumber
\end{eqnarray}
where $\rho_0$ is the standard concentration ($\rho_0=1\,$M$\cdot$liter$^{-1}$) while $\Omega_{ij}(\{ {\bf r} \})$ and $\Omega_{i}^\mathrm{s} (\{ {\bf r} \}) $ are the volume of the configurational space available to each linkage cross-linking particle $i$ with $j$, and particle $i$ with the surface, respectively. Note how configurational contributions depend on the position of the particles $\{ {\bf r} \}$. In this work we consider reactive sequences tethered to particles' surfaces through short, thin rods of double-stranded DNA of length $L$ (see Fig.~\ref{Fig:System}). When $L$ is much smaller than the radius of the particles, the reactive sequences of unbound linkers are uniformly distributed within the layer of thickness $L$ surrounding the tethering surfaces \cite{Angioletti-Uberti_PRL_2014}. In this limit, the configurational space of bound sequences ($\Omega_{ij}$ and $\Omega^\mthr{s}_i$) is the volume of the overlapping regions spanned by the reacting sequences before binding (see Fig.~\ref{Fig:System}$b$). Similarly, the volumes available to unbound reactive  sequences ($\Omega^\mthr{s}$ and $\Omega_i$) are depleted by the volume excluded by the hard-core of the neighboring particles (respectively, $k^\mthr{s}_i$ and $e_{ij}$ in Fig.~\ref{Fig:System}$b$) and of the surface ($e^\mthr{s}_i$ in Fig.~\ref{Fig:System}$b$). We report the explicit expression of the terms appearing in Eqs.~\ref{Equ:DG} in SI Sec.~1.
When written in term of the stationary number of linkages, the multivalent free energy simplifies into a portable expression $F$ ($F={\cal F}_\mathrm{multi}(\{ \on \})$) given by (see SI Sec.~1) \cite{Angioletti-Uberti_PCCP_2016,Angioletti-Uberti_PRL_2014,Petitzon_SoftMatt_2016}
\begin{eqnarray}
\beta F (\{ {\bf r} \})&=& \sum_{i=1}^{N_\mathrm{p}} \left( N_\mthr{L} \log \frac{ \on^\mthr{A}_i \on^\mthr{B}_i }{ N_\mthr{L}^2} + \on^\mthr{AB}_{ii}  + \on^\mthr{AC}_i \right) +F_{T=\infty}
\label{Equ:FreeMulti} \\
&+&  \sum_{1\leq j < q \leq N_\mthr{p}} (\on^\mthr{AB}_{jq}+ \on^\mthr{BA}_{jq})+ N_\mthr{C} \log \frac{ \on^\mthr{C}_\mthr{s} }{ N_\mthr{C}} \, ,
\nonumber
\end{eqnarray}
where $F_{T=\infty}$ is the free energy without any linkage (as found at infinite temperature, $T$) and accounts for non-selective interactions and repulsive osmotic terms due to compression of the linkers in the contact region \cite{Dreyfus_PRL_2009,Angioletti-Uberti_PRL_2014,Angioletti-Uberti_PCCP_2016}. We use Eq.~\ref{Equ:FreeMulti} to sample colloidal configurations $\{ {\bf r} \}$ in the Mean Field calculations (Sec.~\ref{Sec:Programming}) and in simulations (Sec.~\ref{Sec:Sim}).

\subsection{Simulation methods}\label{Sec:Sim}
Modelling self-assembly dynamics of particles forming reversible linkages requires an algorithm capable of evolving the number of linkages $\{ n\}$ and colloids' positions $\{ {\bf r} \}$ in a concerted way \cite{Angioletti-Uberti_PRL_2014,Petitzon_SoftMatt_2016}. At each step of our simulation scheme, we first upgrade the number of linkages between particles, $\{ n\} \to \{n'\}$, while keeping $\{\bf r\}$ fixed. We then calculate forces acting on each particle ${\bf f}_i$ using ${\cal F}_\mathrm{multi}$ (Eq.~\ref{Equ:Z0})  
\begin{eqnarray}
\beta {\bf f}_i &=& -{\bf \nabla}_{{\bf r}_i} \beta {\cal F}_\mthr{multi}(\{n\} , \{ {\bf{r}} \}) 
\label{Equ:force0} \\
&=&  - \frac{\partial }{ \partial \{ \Delta G \} } \beta {\cal F}_\mthr{multi}(\{n\} , \{ \Delta G \}) \cdot {\bf \nabla}_{{\bf r}_i} \{ \Delta G \} \nonumber 
\label{Equ:force}
\end{eqnarray}
where $\{ \Delta G \}$ is the ensemble of possible hybridization free energies (see Eqs.~\ref{Equ:DG}). We provide the explicit expression of ${\bf f}_i$ and further details on the simulation method in SI Sec.~3.
Using ${\bf f}_i$ we evolve particles' position using a Brownian dynamics scheme, $\{{\bf r}'\}=\{{\bf r}\} + \{  \Delta {\bf r} \}$, with
\begin{eqnarray}
\Delta {\bf r}_i &=& {\bf{r}}_i(t+\Delta t) - {\bf{r}}_i(t) = {\bf f}_i \frac{D }{ k_\mathrm{B}T} \Delta t + \sqrt{D \Delta t} {\bf {\cal N}} (0,1) \, ,
\label{Equ:BD}
\end{eqnarray}
where $D$ is the particle diffusion constant, $\Delta t$ the integration step, and ${\bf {\cal N}}$ a normal distributed vector with covariance matrix equal to the unitary matrix. 
\\
We employ two schemes upgrading $\{ n\}$ in different ways. In the implicit scheme (IMP), $\{ n' \}$ are taken equal to the solutions of Eqs.\ \ref{Equ:Saddle} (see SI Sec.~3 for the iterative procedure used to solve Eqs.~\ref{Equ:Saddle} \cite{Angioletti-Uberti_PRL_2014,Petitzon_SoftMatt_2016}). Such scheme minimizes the multivalent free energy ${\cal F}_\mthr{multi}$ at each step of the dynamics, implicitly assuming infinite reaction rates between DNA linkers ($k_\mthr{on}$ and $k_\mthr{off}$ in Fig.~\ref{Fig:System}$c$).
To probe the effect of finite reaction rates on the dynamics of self-assembly, we also developed a scheme in which we explicitly simulate linkages' dynamics using the Gillespie algorithm (EXP) \cite{gillespie1977exact,Petitzon_SoftMatt_2016}.
At a given $\{\bf{r}\}$, we start by calculating the rates at which inter--particle linkages, loops, and particle-surface bridges form
\begin{eqnarray}
 k_\mathrm{on}^{ij}&=&\frac{\Omega_{ij}( \{  {\bf{r}}   \} ) k_\mathrm{on}^0 }{ \Omega_i (    \{  {\bf{r}}  \}  )\Omega_j(\{    {\bf{r}}     \})} 
 \nonumber \\
 k_\mathrm{on}^{ii} &=& \frac{k_\mathrm{on}^0 }{ \Omega_i(\{    {\bf{r}}     \})} 
 \nonumber \\
 k_\mathrm{on}^{i\mathrm{s}} &=& \frac{\Omega_{i}^\mathrm{s}( \{  {\bf{r}}   \} )  k_\mathrm{on}^0 }{ \Omega_i (    \{  {\bf{r}}  \}  )\Omega^\mathrm{s}(\{{\bf{r}}\})}   
\label{Eq:onrates}
\end{eqnarray}
where $k_\mathrm{on}^0$ is the dimerization rate of free oligomers in solutions.  The previous equations follow from the definitions of the hybridization free--energies (Eqs.~\ref{Equ:DG}) while assuming that the $off$ rates of tethered DNA match the $off$ rates of free oligomers in solution \cite{Parolini_ACSNano_2016,Petitzon_SoftMatt_2016,Ho_BiophJ_2009}:
$ k_\mathrm{off}^{ij}= k_\mathrm{off}^{ii}=\rho_0\exp(\beta \Delta G_0) k_\mathrm{on}^0$ 
and $k_\mathrm{off}^{i\mathrm{s}}=\rho_0\exp(\beta \Delta G_0^\mathrm{s}) k_\mathrm{on}^0$.
Once all $on$ and $off$ rates are known, we use the standard procedure of the Gillespie algorithm and sequentially sample one within all possible reactions along with the expected time ($\tau$) for it to happen. We increment a reaction clock ($\tau_\mathrm{reac}$) by $\tau$, $\tau_\mathrm{reac}=\tau_\mathrm{reac}+\tau$ and repeat the procedure until $\tau_\mathrm{reac}$ remains smaller than $\Delta t$. At that point, the Gillespie algorithm is arrested and a Brownian Dynamics step is performed (Eq.~\ref{Equ:BD}). Notice that in our scheme oligomers of the same type on different particles are treated as different reacting species given the fact that the $on$ rates (Eq.~\ref{Eq:onrates}) are configuration dependent. We report further details on the Gillespie algorithm in SI Sec.~3. 
\\
In addition to Brownian dynamics simulations, we perform grand canonical moves allowing to use small simulation boxes without depleting the gas phase (see SI Sec.~3 for details).
\\
Chosen unit of length and time are $L$ and ${L^2 \cdot D^{-1}}$, where $D$ is the diffusion constant of diluted colloids. In particular, the $on$ rate of free reactive sequences in solution ($k_\mathrm{on}^{0}$) is expressed in unit of $LD$ while the hybridization free energies $\Delta G_0$ and $\Delta G^\mthr{s}_0$ have been offset by a constant term equal to $k_B T \log (\rho_0 L^3)$ (see Eqs.~\ref{Equ:DG}).

\section{Results}
\subsection{Programming multibody interactions using mobile ligands}\label{Sec:Programming}
Fig.~\ref{Fig:System}$a$ shows how the functionalized surface alters particle-particle interactions. 
We use Eq.~\ref{Equ:FreeMulti} to calculate the effective interaction $f_{12}$ between pairs of particles in bulk at different particle-particle distance $D$ (Case I) and in contact with the surface (Case II) or with a surface bound colloid (Case II). We offset $f_{12}$ by the value of $F$ at large $D$, $f_{12}= F(D) - F(D=\infty)$ (note that for isolated particles $F=F_\mathrm{multi}< 0$ given that $\on^\mthr{AB}_{ii} > 0$, see Eq.~\ref{Equ:FreeMulti}). In bulk, inter--particle loops dominate (Case I in Fig.~\ref{Fig:System}$a$) resulting in weak particle-particle interactions $f_{12}$ (see full line in the graph of Fig.~\ref{Fig:System}$a$) and a stable gas phase \cite{Bachmann_SoftMatt_2016}. Because receptors (C) at the surface are not self--protected, particles will likely form particle--surface bridges at high enough receptor concentration ($\sigma_\mthr{R}$) and strength (i.e., at low $\Delta G^\mthr{s}_0$). When bound to the surface, a particle will free one linker of type B for each particle-surface bridge formed. Free B ligands will then stabilize another particle (see Case II in Fig.~\ref{Fig:System}$a$). This second binding, as well as the attachment of the colloid to the surface, is thermodynamically more favorable than pairing two particles in bulk because it requires denaturating a single loop (instead of two) to form two inter--particle bridges. This is proven by the graph in Fig.~\ref{Fig:System}$a$ showing how $f_{12}$ is magnified by a factor of ten when one colloid is in direct contact with the surface (see Case I and Case II). The graph of Fig.~\ref{Fig:System}$a$ also shows how, when colloids are not in direct contact with the surface (Case III), effective interactions are much weaker. On the other hand, effective interactions between two surface--bound particles (not shown in Fig.~\ref{Fig:System}a) are weaker than in bulk given the fact that both particles express free ligands of the same type. The change of the state of the ligands displayed by particles following the encounter with the surface underlies the possibility of controlling layer deposition.

\begin{figure}[ht!]
\vspace{0.5cm}
\centering
\includegraphics[width=8.5cm]{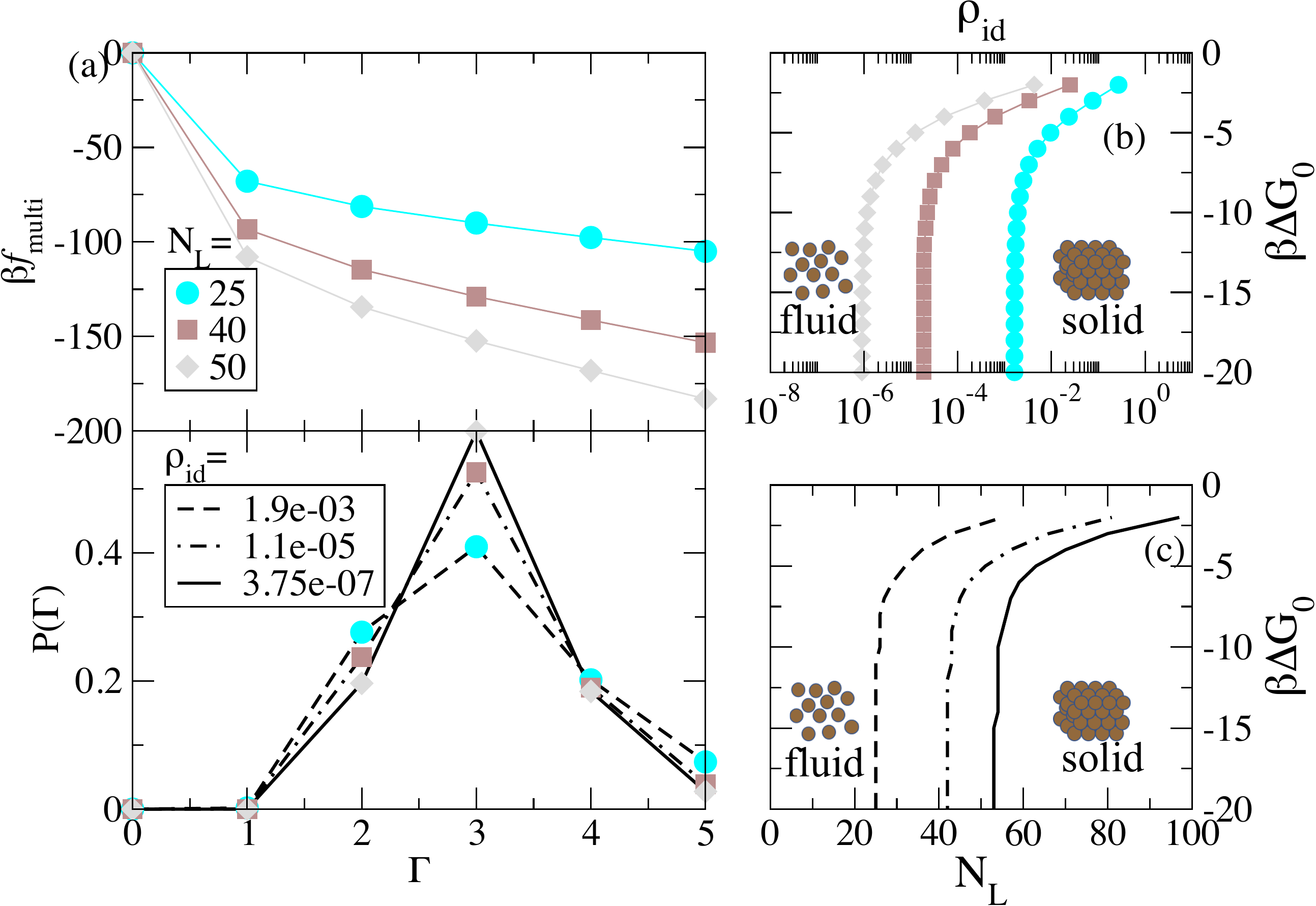} 
%\vspace{-0.5cm}
\caption{ A Mean Field Theory clarifies the parameters controlling the thickness of the crystals. ($a$)~{\it Top panel}: free energy gain of forming crystals comprising $\Gamma$ layers for three different $N_\mthr{L}$. {\it Bottom panel}: Thickness probability for three values of $N_\mthr{L}$ (see legend of the top panel) and three different gas densities. The receptor density and strength are $\sigma_R=1.8\cdot L^{-2}$ and $\beta \Delta G^\mthr{s}_0=-10$, while $\rho_\mthr{id}$ is in units of $L^{-3}$. ($b$), ($c$)~Phase boundaries between the gas and the solid phase in bulk at different $\Delta G_0$. In panel ($b$) we fix $N_\mathrm{L}$ (see top legend of panel $a$) and report the transition density. In panel ($c$) $\rho_{id}$ is fixed (see bottom legend of panel $a$) and we change $N_\mthr{L}$. The MFT uses a fixed particle-particle and particle-surface distance equal to 11$\cdot L$ and 5.7$\cdot L$. \label{Fig:PandBulk} }
\end{figure}
\begin{figure}[ht!]
\vspace{0.5cm}
\centering
\includegraphics[width=6.5cm]{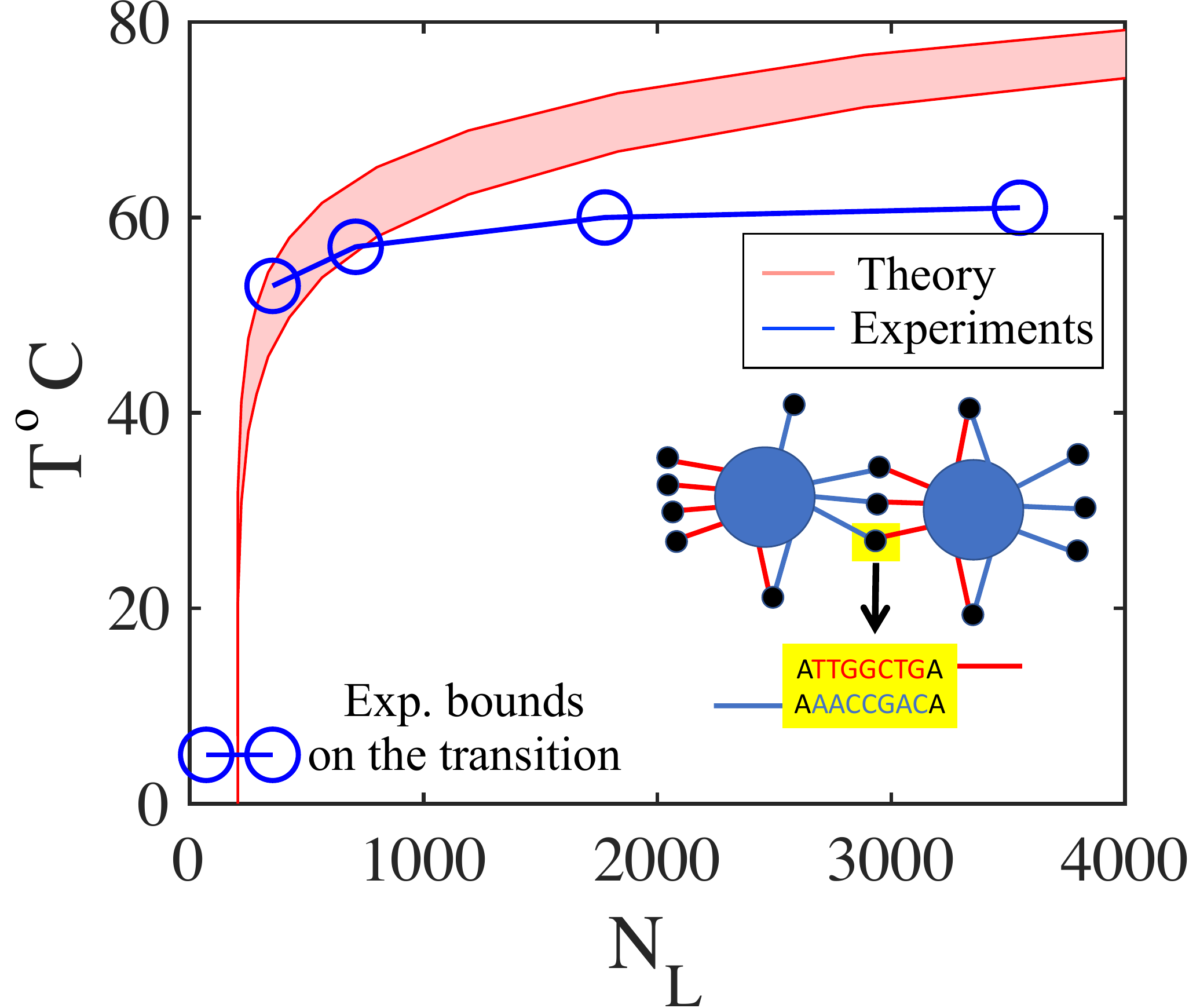} 
%\vspace{-0.5cm}
\caption{Phase boundary of suspensions of 400\,nm diameter vesicles functionalized by complementary linkers of length $L=10\,$nm \cite{Bachmann_SoftMatt_2016}. At low temperature (low $\Delta G_0$), the MFT transition point is located at $N_L \approx 210$ consistent with the experimental result $71< N_L <355$. We use $\rho_\mathrm{id}=2.1\cdot 10^{-9}\,$nm$^{-3}$ \cite{Bachmann_SoftMatt_2016}, and neglect vesicle deformability. We estimate $\Delta G_0$ using nearest neighbor rules \cite{SantaLucia_PNAS_1998} for the reactive sequences reported in the inset. We further decrease $\Delta G_0$ by $2.2\,$Kcal$\cdot$M$^{-1}$ given that Ref.~\cite{Bachmann_SoftMatt_2016} used fluorophores that are known to stabilize the duplex \cite{MOREIRA201536}. The stained region accounts for inaccuracies on the experimental value of $\Delta G_0$ due to the presence of inert tails \cite{Di-Michele_JACS_2014}. An estimation of the magnitude of the tail effect is usually obtained including or not the nearest neighbor contributions of the dangling bases (A bases in black) \cite{Di-Michele_JACS_2014}. \label{Fig:SoftMatt2016} }
\end{figure}
\subsection{Self-assembly of crystals with a finite thickness}

We consider suspensions of colloids as in Fig.~\ref{Fig:System}$a$ and report the formation of crystals at the surfaces. In particular, despite the weak in--plane interactions between particles, we never observe the formation of non--compact structures like colloidal chains. Here, we clarify the system parameters and the thermodynamic conditions controlling the morphology of the assemblies. To do so, we employ the results of a Mean-Field Theory (MFT) balancing the free energy gains of forming an aggregate due to multivalent interactions (Eq.~\ref{Equ:FreeMulti}) with the entropic losses of caging particles into crystalline sites. Using simulations, below we prove that the proposed MFT is quantitative, therefore providing a predictive tool that will be useful to design future experiments. In particular, we used the MFT to fine tune the system parameters of all simulations presented in this work. The SI Sec.~2 reports details of the MFT calculations. The scripts implementing the MFT can be found at \cite{MFTscript} under an MIT license.
\\
We consider the thermodynamic equilibrium between particles in bulk at density $\rho_\mathrm{id}$ and crystalline structures made of $\Gamma$ layers with $N_\mathrm{p,ads}$ particles {\it per} layer ($N_\mathrm{p,ads}=N_\mathrm{ads}/\Gamma$, where $\Gamma$ is the number of layers and $N_\mthr{ads}$ the number of particles in the crystal) for which we calculate the free energy $F(\Gamma)$ using Eq.~\ref{Equ:FreeMulti}. In all cases studied using simulations, we report self-assembly of fcc (111) crystals comprised of hexagonal, stacked layers parallel to the surface. In the top panel of Fig.~\ref{Fig:PandBulk}$a$, we then calculate the multivalent free energy gain {\it per} particle in direct contact with the substrate defined as
\begin{eqnarray}
f_\mathrm{multi}(\Gamma) &=& F(\Gamma)/N_\mathrm{p,ads}-F_1 \Gamma \, ,
\label{Equ:fmulti}
\end{eqnarray}
where $F_1$ is the free energy of a single particle in bulk. We find that $f_\mthr{multi}(\Gamma)$ is non-linear at small $\Gamma$, corresponding to magnified interactions between colloids closer to the functionalized surface (see the graph of $f_{12}$ in Fig.~\ref{Fig:System}$a$). Such nonlinearity is more prominent at high values of $N_\mthr{L}$. At larger values of $\Gamma$, $f_\mthr{multi}(\Gamma)$ becomes linear and surface effects negligible. 
\\
At a given ligand/receptor strength and coating densities, the chemical potential of the colloid, $\mu$, controls the number of layers assembled. In diluted conditions, $\mu$ is proportional to the logarithm of the gas density, $\beta \mu\sim\log \rho_\mthr{id}$. If we assume that the configurational space available to colloids in the crystal phase is $v_0$, the probability of forming crystals made of $\Gamma$ layers is then given by (see SI Sec.~2) \cite{sear1999stability,charbonneau2007gas}
\begin{eqnarray}
P(\Gamma) &=& \frac{1}{Z} (\rho_\mthr{id} v_0)^\Gamma \exp[-\beta  f_\mathrm{multi}(\Gamma)]\, ,
\label{Equ:MFT}
\end{eqnarray}
where $Z$ is a normalization factor. In the bottom panel of Fig.~\ref{Fig:PandBulk}$a$, we report $P(\Gamma)$ for three different values of $N_\mthr{L}$. We choose three different $\rho_\mthr{id}$ resulting in a most likely number of layers equal to three. We verify the predictivity of $P$ using simulations (see below). Note that the definition of $P$ is conditional on having a finite normalization factor $Z$. The onset in parameter space at which $Z$ diverges corresponds to the fluid--solid phase boundary in bulk (see SI Sec.~2) \cite{sear1999stability,charbonneau2007gas}. In Figs.~\ref{Fig:PandBulk}$b$ and $c$ we then report the predicted bulk phase diagram in the $(\Delta G_0,\,\rho_\mthr{id})$ and $(\Delta G_0,\,N_\mthr{L})$ planes. The gas phase is stable at low $N_\mthr{L}$ and $\rho_\mthr{id}$ values. Note how at low $\Delta G_0$ (equivalently, at low temperature) the phase boundary does not depend on $\Delta G_0$. In this limit, the numbers of paired linkers in the gas and crystalline phase are equal, and combinatorial terms fully control the transition \cite{Zilman_PRL_2003,Bozorgui_PRL_2008}. 
 Ref.~\cite{Bachmann_SoftMatt_2016} studied experimentally the self-assembly of 200$\,$nm diameter vesicles functionalized by two families of linkers as in Fig.~\ref{Fig:System}. In Fig.~\ref{Fig:SoftMatt2016}, we adapt the MFT developed in the present work to the system of Ref.~\cite{Bachmann_SoftMatt_2016} and report a phase diagram similar to the one of Fig.~\ref{Fig:PandBulk}$c$. Our MFT allows predicting the entropic transition at low $T$ (low $\Delta G_0$), as well the phase boundary at high temperature, without any fitting parameter (see caption of Fig.~\ref{Fig:SoftMatt2016} for details). The discrepancy between the theoretical and the experimental phase boundary at high $N_L$ (not relevant to this study) are due to steric interactions between linkers not included in our MFT \cite{Bachmann_SoftMatt_2016}. Overall, Fig.~\ref{Fig:SoftMatt2016} validates our modeling and suggests using (supported) vesicles in future experiments aiming at reproducing controllable colloidal layer deposition.  
\begin{figure}[ht!]
\vspace{0.5cm}
\centering
\includegraphics[width=8.5cm]{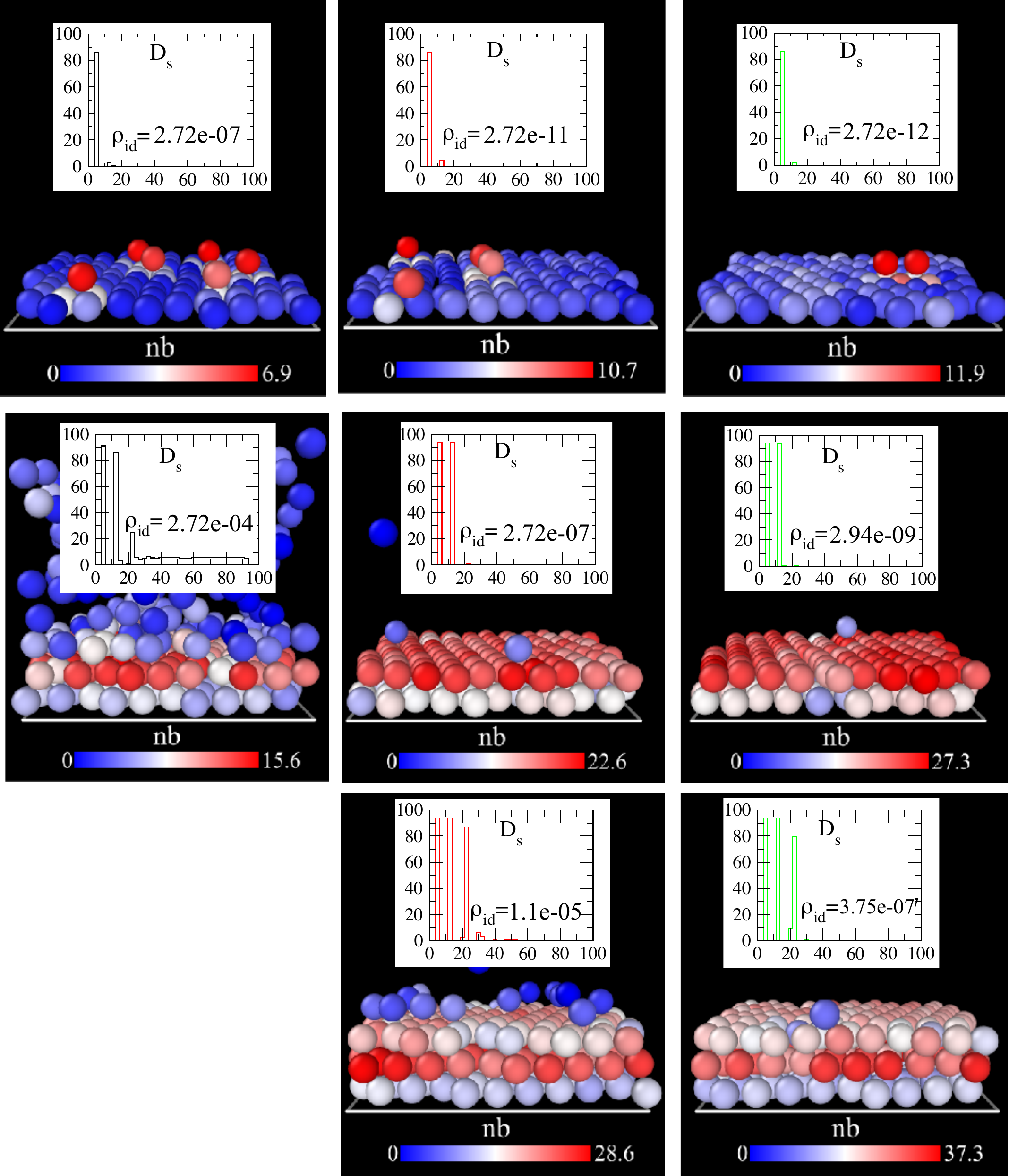} 
%\vspace{-0.5cm}
\caption{Brownian Dynamics simulations lead to self-assembly of fcc (111) crystals with controllable thickness. For $N_\mathrm{L}=25$ (first column), $N_\mthr{L}=40$ (second column), and $N_\mthr{L}=50$ (third column) we fine--tune the density of the gas phase using the MFT (Eq.~\ref{Equ:MFT}) and prove the ability of our system to assemble crystals comprising a sought number of layers. The colormap provides the number of particle--particle linkages (nb) and shows how, in the top layer, particles feature fewer bridges. We failed to identify three layers for the $N_\mathrm{L}=50$ system (not shown) due to the high density of the gas phase. Supplementary videos 1, 2, and 3 show simulation trajectories leading, respectively, to the top, middle, and bottom configurations of the $N_\mthr{L}=40$ system. The receptor density is equal to $\sigma_R=1.8\cdot L^{-2}$ corresponding to 11.3, 7.1, and 5.7 times the total densities of ligands on particles with, respectively, $N_\mthr{L}=25$, 40, and 50. \label{Fig:MainSim} }
\end{figure}
\\
We now relax the approximations employed by the MFT and develop Brownian dynamics simulations. For three different values of $N_\mthr{L}$, we run simulations at different gas densities $\rho_\mthr{id}$ and verify the possibility of assembling crystals made of $\Gamma=1$, 2, and 3 layers. Fig.~\ref{Fig:MainSim} reports snapshots obtained in steady conditions. The choice of $\rho_\mthr{id}$ at different $N_\mthr{L}$ has been guided by the MFT providing predictions of the averaged number of layers, $\langle \Gamma \rangle$, through Eq.~\ref{Equ:MFT}. As anticipated before, most of our simulations reported the formation of fcc (111) crystals. Occasionally we observed fcc (100) crystals. Crystals of this type are arrested states, occasionally appearing when using small simulation boxes and high rates of particle insertion/deletion in the grand-canonical scheme. In particular, at high insertion rates, the second layer starts to form before the first layer could relax into a triangular lattice. 
\\
Given that our design is based on equilibrium considerations (see Eq.~\ref{Equ:MFT}), different initial conditions lead to the same number of layers. Supplementary videos 1, 2, and 3 prove that the configurations reported in Fig.~\ref{Fig:MainSim} are stationary. The insets of the panels in Fig.~\ref{Fig:MainSim} report the histograms with the number of particles found at different distances $D_\mthr{s}$ (expressed in units of $L$) from the plane averaged over steady configurations. Such histogram sharply transits from the maximum number of particles that can be fitted into a single plane to zero at the solid-fluid interface. For the system employing the smaller value of $N_\mthr{L}$ the gas phase is denser because lower values of $N_\mthr{L}$ result in higher $f_\mthr{multi}$ (Eq.~\ref{Equ:fmulti}), and higher $\rho_\mthr{id}$ are required to yield a given number of layer $\Gamma$ (see Eq.~\ref{Equ:MFT}). Overall, Fig.~\ref{Fig:MainSim} nicely demonstrates how the proposed system can be used to fabricate crystals with a prescribed number of layers without any direct intervention.

\begin{figure}[ht!]
\vspace{0.5cm}
\centering
\includegraphics[width=8.4cm]{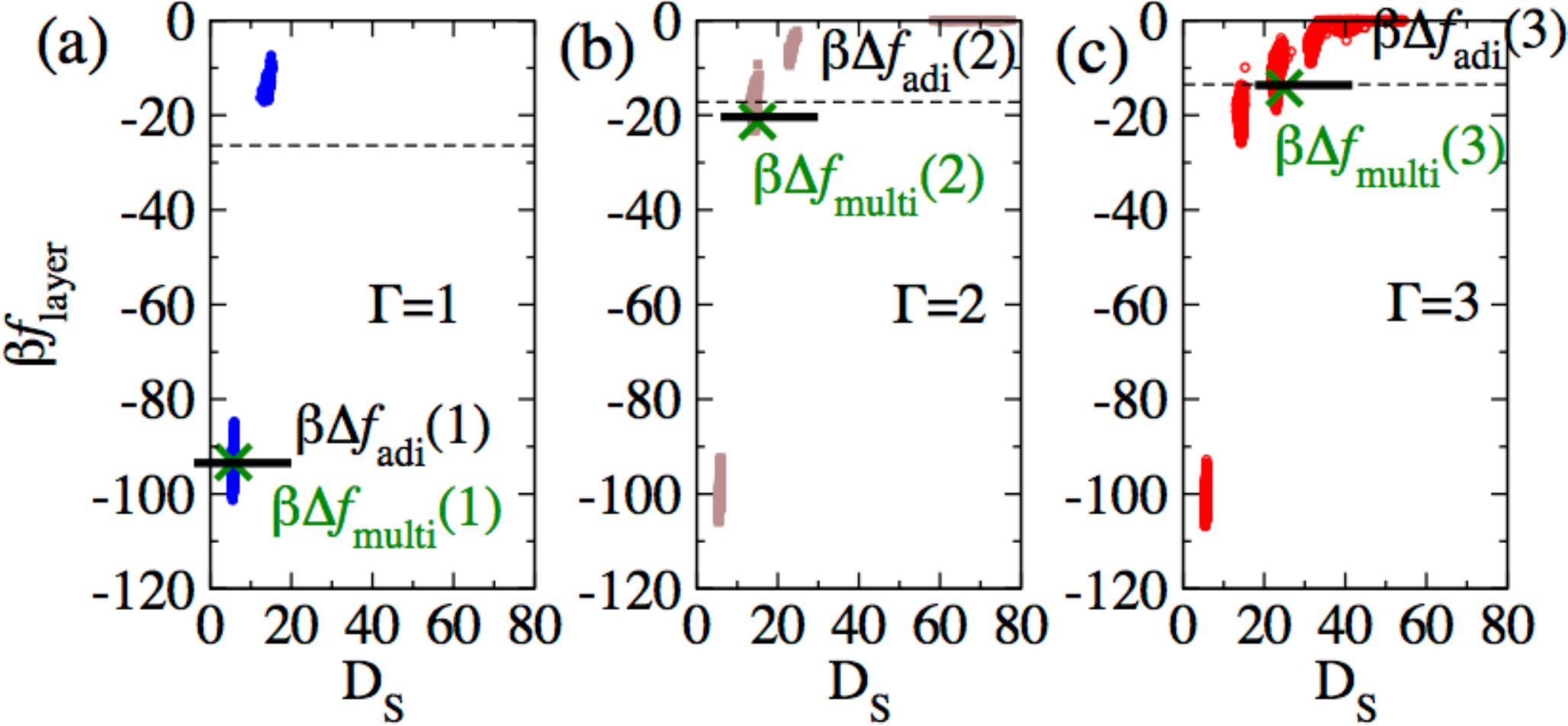} 
%\vspace{-0.5cm}
\caption{Single layer contributions to the multivalent free energy in systems with $N_\mthr{L}=40$ forming $\Gamma=1$, 2, and 3 layers (panel $a$, $b$, and $c$). The stained regions correspond to values of the free energy {\em per} particle ($f_\mthr{layer}$ defined in the text) at a given particle--plane distance $D_\mthr{s}$ sampled using Brownian Dynamics simulations. For comparison, the green X marks refer to the MFT predictions of the free energy gain of adding the top layer, $\Delta f_\mthr{multi}(\Gamma)$, while the horizontal bars are analytic approximations, $\Delta f_\mthr{adi}(\Gamma)$. The dashed lines correspond to $\log (\rho_\mthr{id} v_0)$ (see Eq.~\ref{Equ:MFT}) with $\rho_\mthr{id}$ as in the second column of Fig.~\ref{Fig:MainSim} and $v_0=(L/2)^3$ (see Methods section). The receptor density is equal to $\sigma_R=1.8\cdot L^{-2}$.
\label{Fig:FreeEnLayers} }
\end{figure}
\begin{figure}[ht!]
\vspace{0.5cm}
\centering
\includegraphics[width=7.5cm]{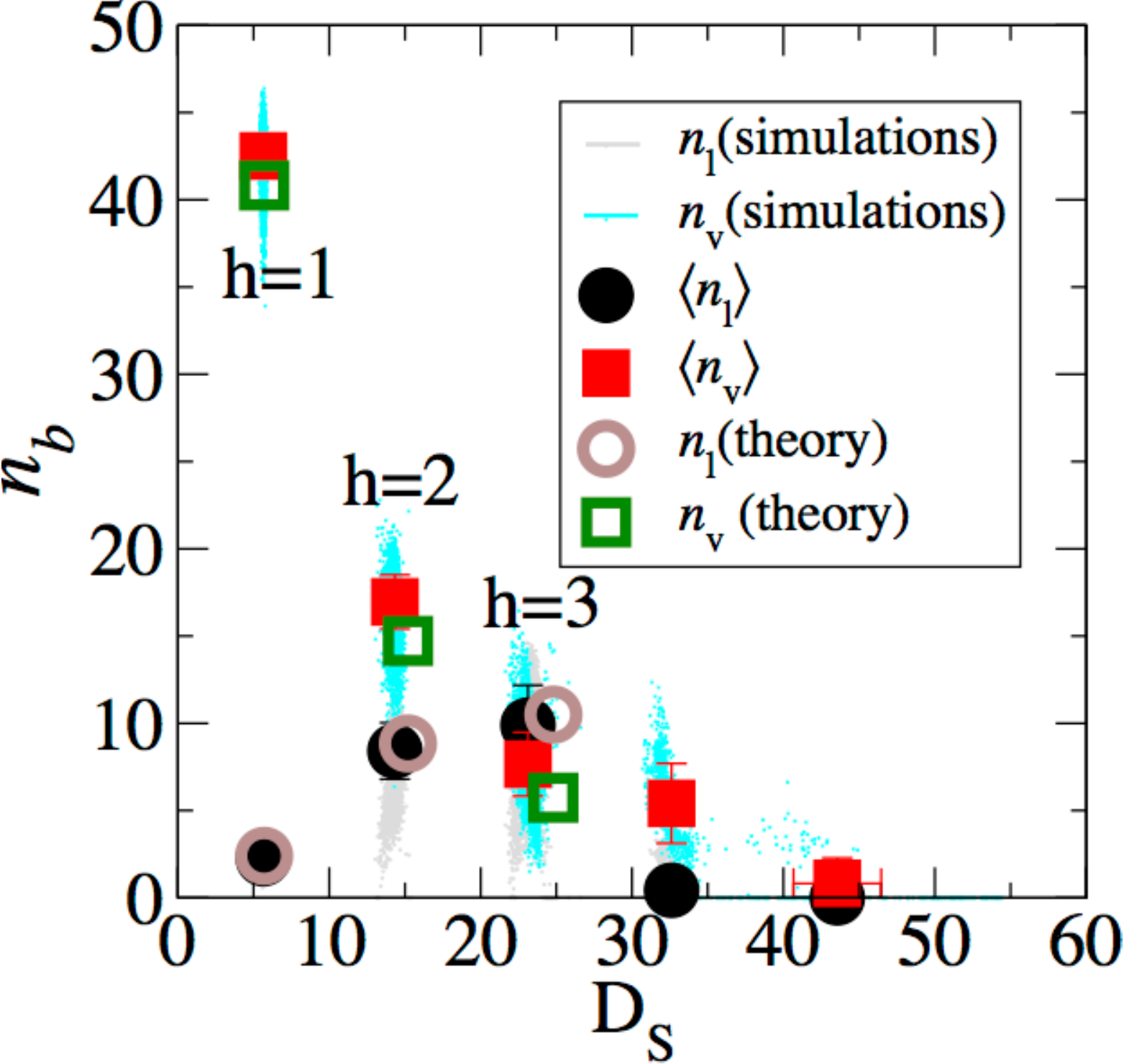} 
%\vspace{-0.5cm}
\caption{Heterogeneity of the number of linkages across the assembled crystals. For the system of Fig.~\ref{Fig:FreeEnLayers}$c$ ($N_\mthr{L}=40$ and $\rho_\mthr{id}=1.1\cdot 10^{-5}$), we report the number of in--plane and off--plane bridges featured by colloids sitting in different layers (tagged with $h$). Close to the surface ($h=1$), off--plane bridges dominate. Stained regions mark values sampled by Brownian Dynamics simulations, full symbols are simulation averages, while empty symbols are MFT predictions. The receptor density is equal to $\sigma_R=1.8\cdot L^{-2}$. \label{Fig:FreeEnLayers-b} }
\end{figure}
\subsection{Controlling thickness in colloidal layer deposition}
In this section, we study in more detail the thermodynamic properties of the structures reported in Fig.~\ref{Fig:MainSim}. 
We consider the system with $N_\mthr{L}=40$ at three different chemical potentials forming from one to three layers as obtained in the second column of Fig.~\ref{Fig:MainSim}.
In Fig.~\ref{Fig:FreeEnLayers} we use simulations to study the contribution ($f_\mthr{layer}$) of particles found at a given layer (specified by the plane-particle distance $D_\mthr{s}$) to the multivalent free energy $F$. Specifically, $f_\mthr{layer}$  is calculated using the values of $\{\on\}$ sampled by the simulations along with Eq.~\ref{Equ:FreeMulti} in which the terms involving $\on^\mthr{AB}_{jq}$ and $\on^\mthr{BA}_{jq}$ are evenly split between particles $j$ and $q$. 
For perfect crystals, we have $f_\mthr{multi}(\Gamma)= \sum_{i=1}^\Gamma f_\mthr{layer}(i)$, with $f_\mthr{multi}$ given by Eq.~\ref{Equ:fmulti}. Consistently with the behavior of $f_{12}$ in Fig.~\ref{Fig:System} and $f_\mthr{multi}$ in Fig.~\ref{Fig:PandBulk}, we find that $f_\mthr{layer}$ sharply increases with the particle-surface distance, $D_\mthr{s}$.
As a result, when considering systems assembling multiple layers, colloids in the
bottom layers are irreversibly adsorbed while at higher layers adsorption/desorption energies
become thermally accessible. As predicted by Eq.~\ref{Equ:MFT}, layers with $f_\mthr{layer}$ greater than $k_B T \log(\rho_\mthr{id}v_0)$ are unstable (see dashed lines in Fig.~\ref{Fig:FreeEnLayers}$a$-$c$).
\\
Although $f_\mthr{layer}(i)$ is also a function of $\Gamma$ (given that the numbers of linkages $\{\on \}$ featured by the particles in the crystal are coupled variables), Fig.~\ref{Fig:FreeEnLayers} shows that such dependency is weak. On the one hand, this observation suggests that $f_\mthr{layer}(\Gamma)$ is well approximated by $\Delta f_\mthr{multi}(\Gamma)\equiv f_\mthr{multi}(\Gamma)- f_\mthr{multi}(\Gamma-1)$ as confirmed by Fig.~\ref{Fig:FreeEnLayers}$a$-$c$. Moreover, Fig.~\ref{Fig:FreeEnLayers} also suggests that adding an extra layer to the crystal has little effect on the number of linkages already formed. This motivated a simplification of the MFT in which, when calculating $\Delta f_\mthr{multi}(\Gamma+1)$, we treat the number of free receptors featured by particles sitting in the layer $\Gamma$ as a constant parameter. In the low--temperature regime (low $\Delta G_0$ in Fig.~\ref{Fig:PandBulk}$b$, $c$), this assumption allows validating an analytic expression, $\Delta f_\mthr{adi}$, that matches well with the values of $\Delta f_\mthr{multi}$ reported in Fig.~\ref{Fig:FreeEnLayers}. The explicit expression of $\Delta f_\mthr{adi}$ and further checks of its validity are reported in SI Sec.~2. In general, we advise using the MFT to refine experimental designs \cite{MFTscript}.
\\
In Fig.~\ref{Fig:FreeEnLayers-b} we study the number of inter--particle bridges featured by colloids found at different layers ($h=1,\, 2$, and 3) of steady configurations of a system assembling a crystal with $N_\mthr{L}=40$ and $<\Gamma>=3$ (corresponding to Fig.~\ref{Fig:FreeEnLayers}$c$). We decompose the total number of bridges emanating from particles belonging to a given layer (tagged with $h$ in Fig.~\ref{Fig:FreeEnLayers-b}) into in--plane, $n_\mthr{l}$, and off--plane linkages, $n_\mthr{v}$. In--plane bridges involve the six neighboring particles of a triangular lattice while off--plane bridges the three+three neighboring particles of an fcc (111) crystal belonging to layers $h\pm 1$ (when present) or the surface if $h=1$. We report simulation data points along with their averages and mean-field predictions. Simulations and theoretical predictions are in agreement. Close to the surface, all bridges are off--plane. This finding shows how two surface--bound particles exhibit an excess of free B ligands and few intra--particle or in--plane bridges. The number of off--plane and in--plane linkages converge when increasing $D_\mthr{s}$, as expected in bulk conditions. However, for the system of Fig.~\ref{Fig:FreeEnLayers-b}, the total number of bridges goes to zero at high $D_\mthr{s}$ in favors of loops consistent with the fact that, in bulk, the gas phase is stable, see Fig.~\ref{Fig:PandBulk}$b$--$c$. Notice how particles (and occasionally dimers) can temporarily bind colloids in the third layer but rapidly detach as confirmed by the fact that they do not form any lateral bridge (see Supplementary video 3). 
\begin{figure}[ht!]
\vspace{0.5cm}
\centering
\includegraphics[width=8.5cm]{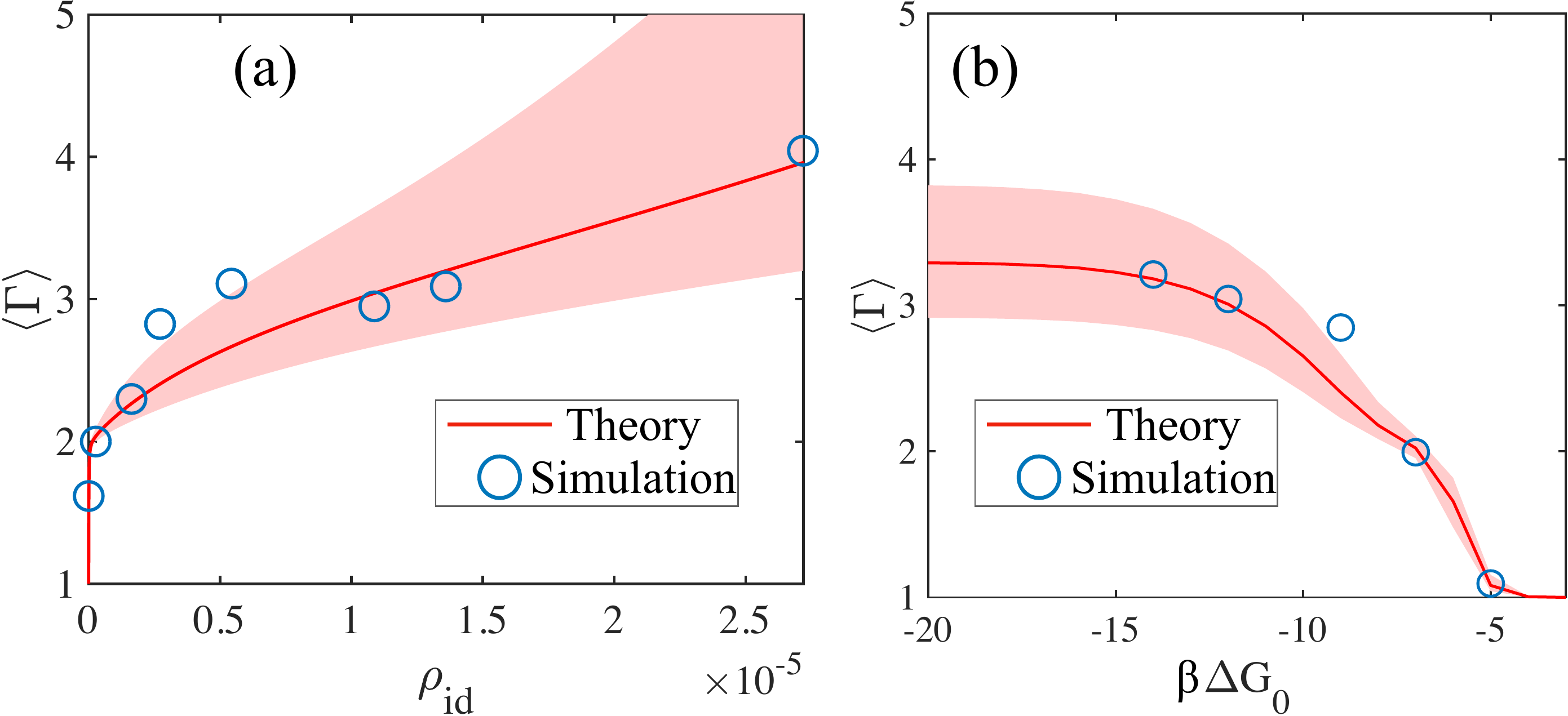} 
%\vspace{-0.5cm}
\caption{The mean field theory predicts the number of layers found in simulations. Averaged number of layers as a function of ($a$) $\rho_\mthr{id}$ (if $\Delta G_0=-9$) and ($b$) $\Delta G_0$ (with $\rho_\mthr{id}=2.72\cdot 10^{-6}$). In both cases $N_\mthr{L}=40$, $\Delta G^\mthr{s}_0=\Delta G_0-1$ and $\sigma_R=1.8 \cdot L^{-2}$. In the MFT, we fixed the configurational volume available to particles in the crystal phase to $v_0=(L/2)^3$. The magenta stained bands refer to theoretical predictions with the value of $v_0$ augmented/decreased by a factor of up to two.\label{Fig:MFTpredictions} }
\end{figure}
\\
In Fig.~\ref{Fig:MFTpredictions} we summarize the number of self-assembled layers as a function of the chemical potential of the system, Fig.~\ref{Fig:MFTpredictions}$a$, and the hybridization free energies of the reactive sequences, Fig.~\ref{Fig:MFTpredictions}$b$. We compare numerical simulations with theoretical predictions and report a quantitative agreement. Compared to theoretical results, numerical predictions do not feature fractional values of $\Gamma$. In particular, in our simulations, we never observe half-filled top layers. SI Fig.~10 reporting the number of adsorbed particles at different chemical potentials versus time confirms this result. Line tension effects, not considered by the MFT, are likely to play a role in hampering the formation of terraced structures. Avoiding terracing is a further advantage of the present design that makes our approach potentially interesting in applications seeking at forming flat interfaces. Fig.~\ref{Fig:MFTpredictions} also investigates the effect of systematic errors of our evaluation of the configurational volume available to colloids in the crystal phase ($v_0$) on the MFT predictions. 
\\
We can use our quantitative MFT for design purposes.  In SI Fig.~11 we extend the analysis of Fig.~\ref{Fig:MFTpredictions} and report an extensive study of the expected number of layers, $\langle \Gamma \rangle$, as a function of $\Delta G_0$, $N_\mthr{L}$, and $\rho_\mthr{id}$. As expected, $\Gamma$ diverges as we approach the bulk phase boundary (see Figs.~\ref{Fig:PandBulk}$b$ and $c$) from the gas phase either by increasing the number of linkers, $N_\mthr{L}$, or the chemical potential, $\rho_\mthr{id}$. Moreover, we can also investigate the effect of changing the strength of particle--surface relative to particle--particle hybridization free energy, $K=\Delta G_0-\Delta G^\mthr{s}_0$. SI Fig.~12 shows how, by increasing $K$, the system sharply transits from not forming any layer (the surface cannot denaturate loops) to a constant thickness (no loops left on the particles of the first layer). SI Figs.~11, 12 clarify how, generally, the averaged number of layers is not a function of $\Delta G_0$. This highlights the fact that layering is solely controlled by entropic factors (namely combinatorial terms and the ratio between the statistical weight of loops and bridges) as already observed for the bulk gas--solid transition (see Fig.~\ref{Fig:PandBulk}$b$ and $c$).

\section{Discussions and Conclusions}

The results of the previous section show the robustness and flexibility of our design strategy to achieve colloidal layer deposition. Our system provides multiple tuning parameters allowing to control the thickness of crystals self-assembled at the functionalized surface. For instance, as shown in Fig.~\ref{Fig:MainSim} and SI Fig.~11 and 12, using the number of linkers {\it per} particle, $N_\mthr{L}$, as a design parameter could elude experimental limitations on the possible values of the chemical potential ($\rho_\mthr{id}\sim \exp [\beta \mu]$). We notice how $N_\mthr{L}$ has been already employed in recent experiments (Ref.~\cite{Bachmann_SoftMatt_2016}) to control aggregation of DNA functionalized vesicles resulting in a phase diagram that we have successfully reproduced in the present study (see Fig.~\ref{Fig:SoftMatt2016}).
\\
The rigorous free-energy calculations presented in the previous sections certify the generality of the mechanism leading to controllable layer deposition. The only two indispensable ingredients underlying the sought effect are the mobility of the binders and the presence of competition between inter-- and intra--particle linkages. Nevertheless, in systems of DNA mediated interactions, kinetic limitations may hamper yielding, for instance, of thermodynamically stable crystals in favor of disordered aggregates. In the last five years, several design strategies have been proposed leading to enhanced crystallization. In particular, Van der Meulen and Leunissen clarified the advantages of using mobile linkers when aiming at yielding regular structures as due to the possibility of bound particles to pivot around each other \cite{Meulen_JACS_2013}. 
\begin{figure}[ht!]
\vspace{0.5cm}
\centering
\includegraphics[width=8.5cm]{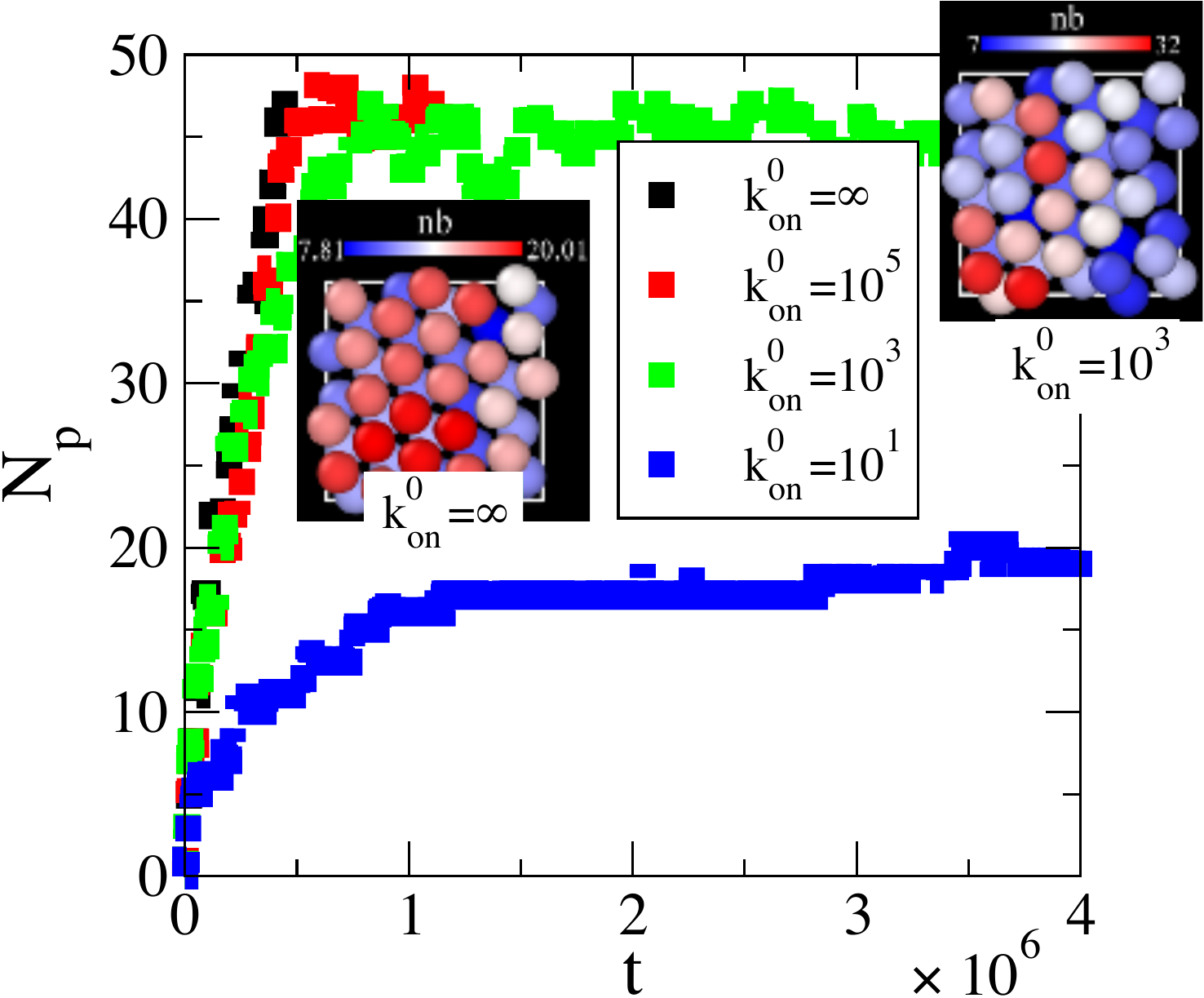} 
%\vspace{-0.5cm}
\caption{DNA reactions rates do not affect the properties of the self-assembled crystal.  (a) Number of adsorbed particles versus time for three different $on$ rates (expressed in units of $D\cdot \rho_0^{-1}\cdot L^{-2}$) as compared to implicit linkage simulations (IMP) assuming $k^0_\mthr{on}=\infty$ (see Methods section). Notice that $k^0_\mthr{on}$ controls both $on$ and $off$ rates ($k_\mthr{on}$ and $k_\mthr{off}$ in Fig.~\ref{Fig:System}$c$) while does not affect the hybridization free energies (see Methods section). Insets show two-layer crystals obtained using $k^0_\mthr{on}=10^3$ (top-right) and IMP simulation (left). The colormap highlights the number of particle-particle bridges ($nb$). \label{Fig:EXP} }
\end{figure}
\\
Beyond defect annihilation, the finite rates at which DNA linkages form and break could affect self-assembly kinetics. We recently showed how, at low temperature, colloidal systems carrying mobile binders struggle to form compact aggregates as due to small DNA denaturation rates impeding re-distribution of interparticle bridges and effectively stabilizing low valency aggregates \cite{Petitzon_SoftMatt_2016}. To probe the effect of finite reaction rates, in Fig.~\ref{Fig:EXP} we report the outcomes of expensive simulations in which DNA reaction dynamics is explicitly simulated using a Gillespie algorithm fed by the rates of forming/breaking DNA linkages (see the EXP method Sec.~\ref{Sec:Sim}). The reaction kinetics is parametrized by the $on$ rate of free reactive oligomers in solution, $k^0_\mthr{on}$, expressed in units of $D\cdot \rho_0^{-1}\cdot L^{-2}$, where $D$ is the diffusion coefficient of diluted colloids and $\rho_0$ the standard concentration. Finite values of $k^0_\mthr{on}$ sensibly slow down adsorption to the point that steady states cannot be reached using affordable simulations. For $k^0_\mthr{on} \to \infty$ we recover the results found using Brownian dynamics simulations calculating the number of linkages using chemical equilibrium equations (IMP in Fig.~\ref{Fig:EXP}). Importantly Fig.~\ref{Fig:EXP}$b$ shows that, in the present system, finite reaction kinetics do not hamper the relaxation of the system toward the equilibrium state. We also notice that it is possible to design suitable DNA sequences capable of converting intra--particle loops into inter--particle bridges without the need of denaturating DNA. In particular, the scheme that has been recently validated by Parolini {\it et al } \cite{Parolini_ACSNano_2016} or by Zhang {\it et al} \cite{zhang2017sequential}, based on the toehold-exchange mechanism of Zhang and Winfree \cite{zhang2009control}, could be readily implemented in the present system to speed up $k^0_\mthr{on}$ by orders of magnitude \cite{zhang2009control}.
\\
In conclusions, in the present contribution, we presented a system capable of self-assembling fcc (111) crystals at functionalized surfaces with finite thickness fully controlled by design and thermodynamic parameters. Beyond addressing an important technological challenge, the proposed system is innovative in that it provides a general design principle to self-assemble finite and localized colloidal aggregates. Often self-assembly of finite structures relies on the use of directional interactions \cite{patra2017layer,halverson2013dna} involving many different particles \cite{Zeravcic_RMPhys_2017,zeravcic2014size}. Such design is experimentally challenging, mainly due to the need of designing many orthogonal pair-interactions with comparable strength. Our strategy provides a valuable alternative to the use of multicomponent systems. Our design principle uses particles that are reprogrammed and interact differently when in contact with functionalized surfaces. We are therefore confident that our work will inspire new investigations and experiments leading to functional materials made of dynamic, trajectory-dependent building blocks.

\section*{Acknowledgements}
This work was supported by the Fonds de la Recherche Scientifique de Belgique - FNRS under grant n$^\circ$ MIS F.4534.17. Computational resources have been provided by the Consortium des \'Equipements de Calcul Intensif (CECI), funded by the Fonds de la Recherche Scientifique de Belgique - FNRS under grant n$^{\circ}$  2.5020.11.

\begin{widetext}

\section*{Supplementary Figures}

\begin{figure}[ht!]
\begin{center}
\includegraphics[width=8cm]{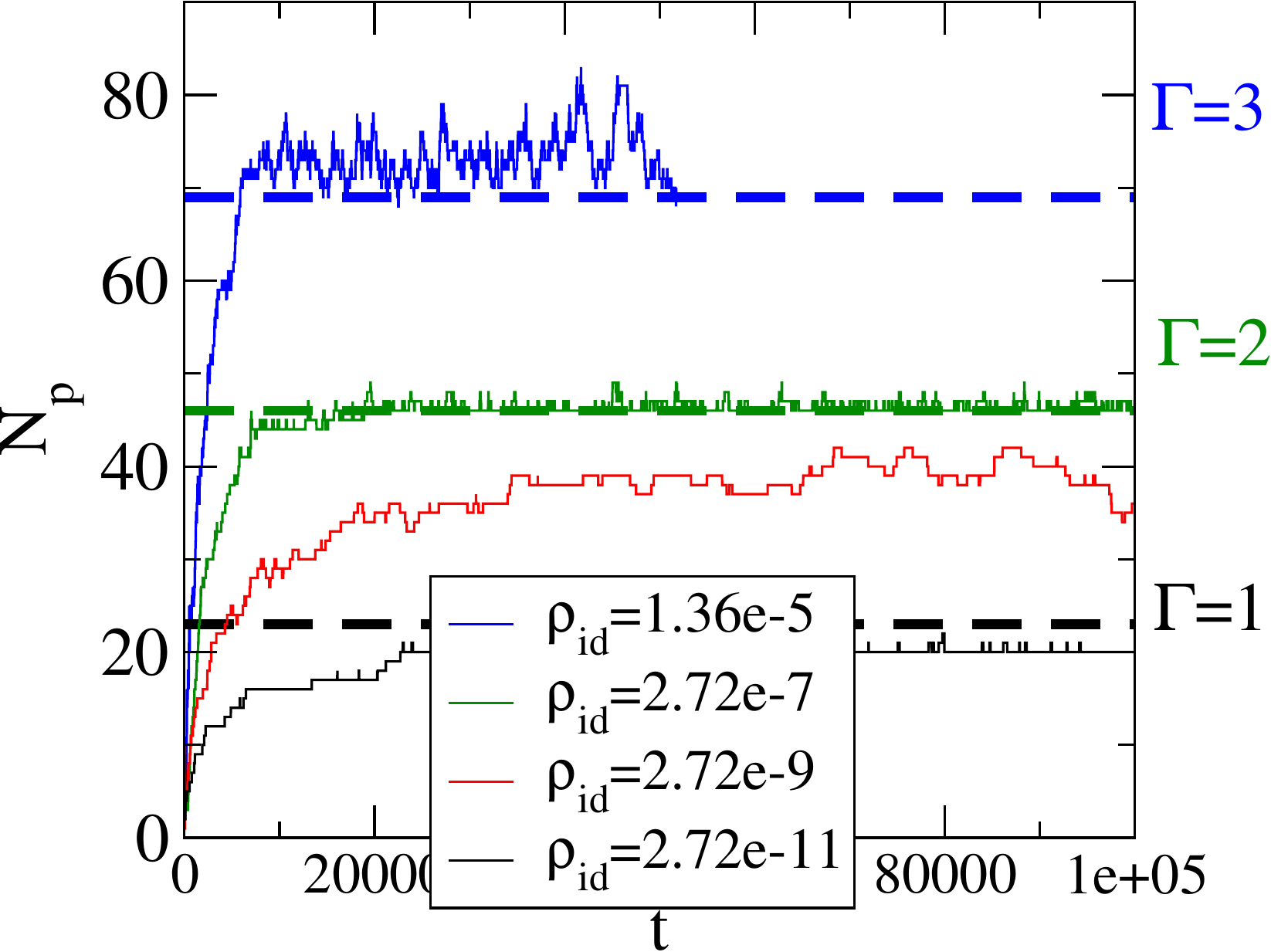}
\caption{\textbf{Self-assembly directed by functionalized interfaces always leads to crystals with filled top layers.} We report the number of adsorbed colloids versus time obtained in simulations using different chemical potentials ($\rho_\mthr{id}\sim\exp[\beta \mu]$). The dashed lines mark the number of colloids compatible with an integer number of layers, $\Gamma$. We used $N_\mthr{L}=40$, $\beta \Delta G_0=\beta \Delta G^\mthr{s}_0+1=-9$, and receptor density $\sigma_R=1.8 \cdot L^{-2}$. }\label{Fig:NoTerraces}
\end{center}
\end{figure}

\begin{figure}[ht!]
\begin{center}
\includegraphics[width=11cm]{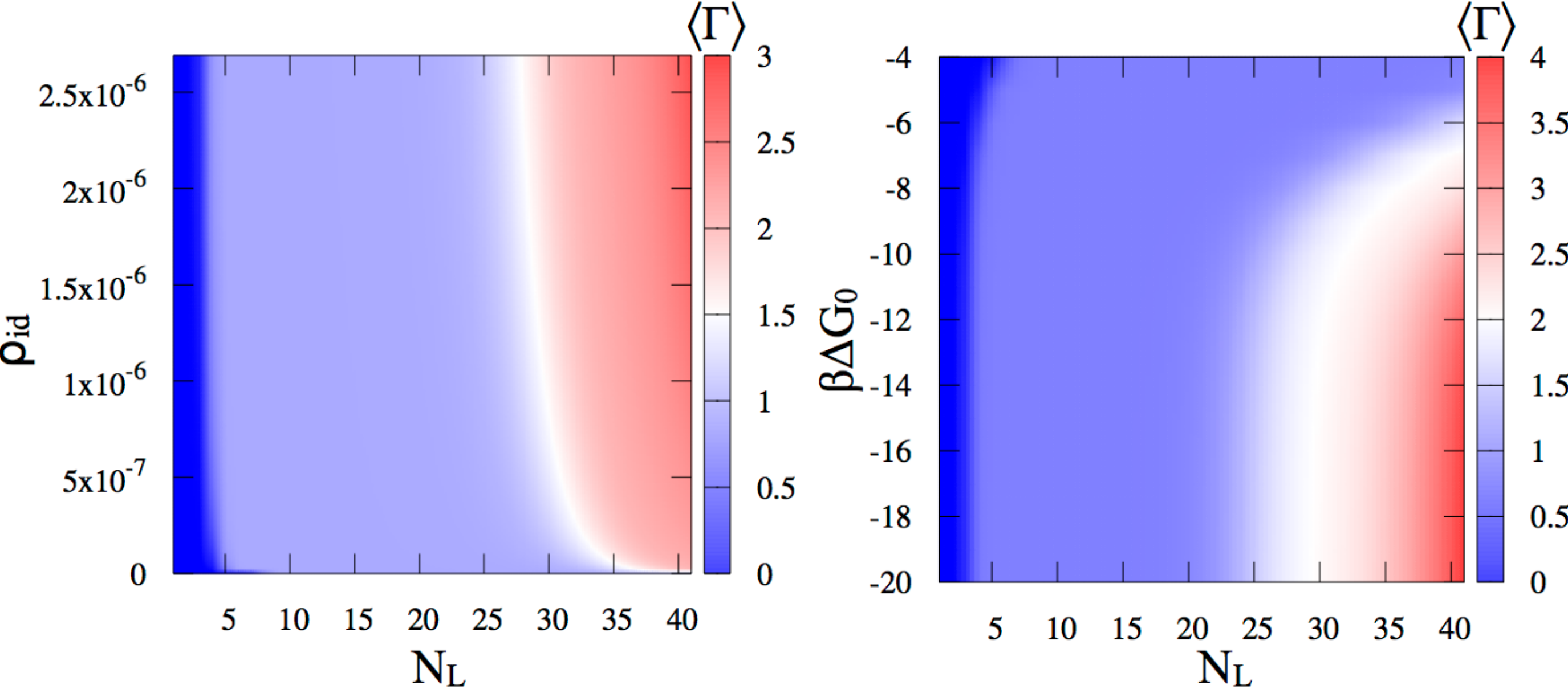}
\caption{\textbf{Mean field theory predictions of the averaged number of self-assembled layers.} Effect of the number of ligands, the density of the gas phase, and ligand-ligand hybridization free-energy on the thickness of the self-assembled crystal. In panel (a) we used $\beta \Delta G_0=-9$ while in panel (b) $\rho_\mthr{id}=2.72 \cdot 10^{-6} L^{-3}$. We used $\beta \Delta G^\mthr{s}_0=\beta \Delta G_0-1$ and receptor density $\sigma_\mthr{R}=1.8\cdot L^{-2}$.  }\label{Fig:AveragedNumbLayers1}
\end{center}
\end{figure}

\begin{figure}[ht!]
\begin{center}
\includegraphics[width=14cm]{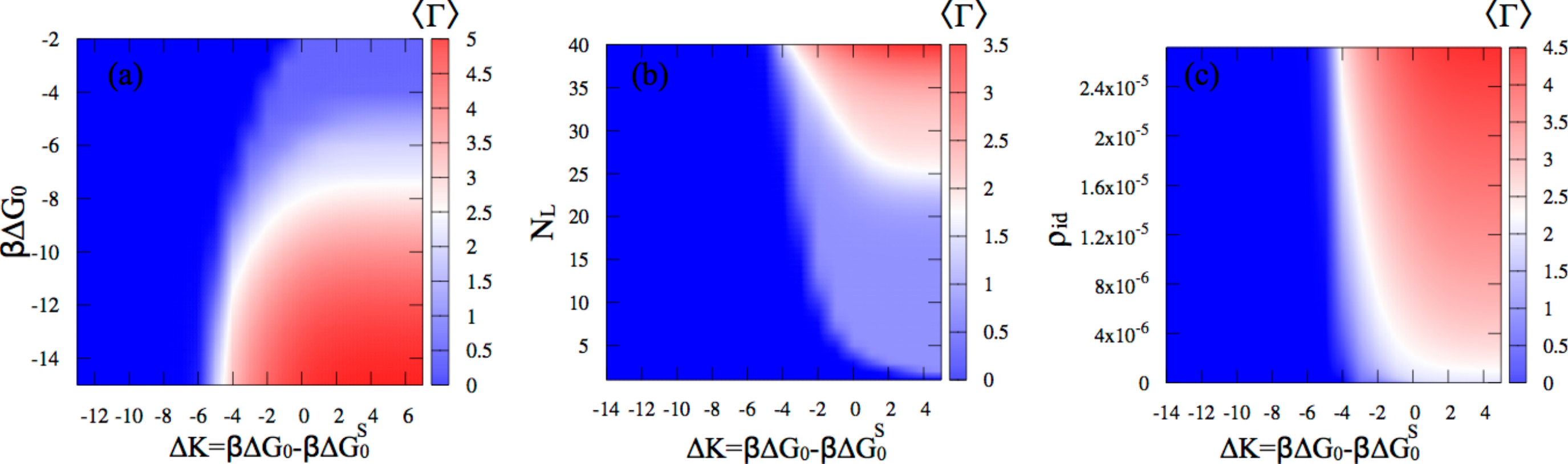}
\caption{\textbf{Effect of the ligand--receptor hybridization free energy on the crystal thickness.} While decreasing $\Delta G^\mthr{s}_0$, the averaged number of layers increases until reaching an asymptotic value. When not varied, we set  $\beta \Delta G_0=-9$, $N_\mthr{L}=40$, and $\rho_\mthr{id}=1.1\cdot 10^{-5} \cdot L^{-3}$. For the receptor density we use $\sigma_\mthr{R}=1.8\cdot L^{-2}$. }\label{Fig:AveragedNumbLayers2}
\end{center}
\end{figure}

\begin{figure}[ht!]
\begin{center}
\includegraphics[width=11cm]{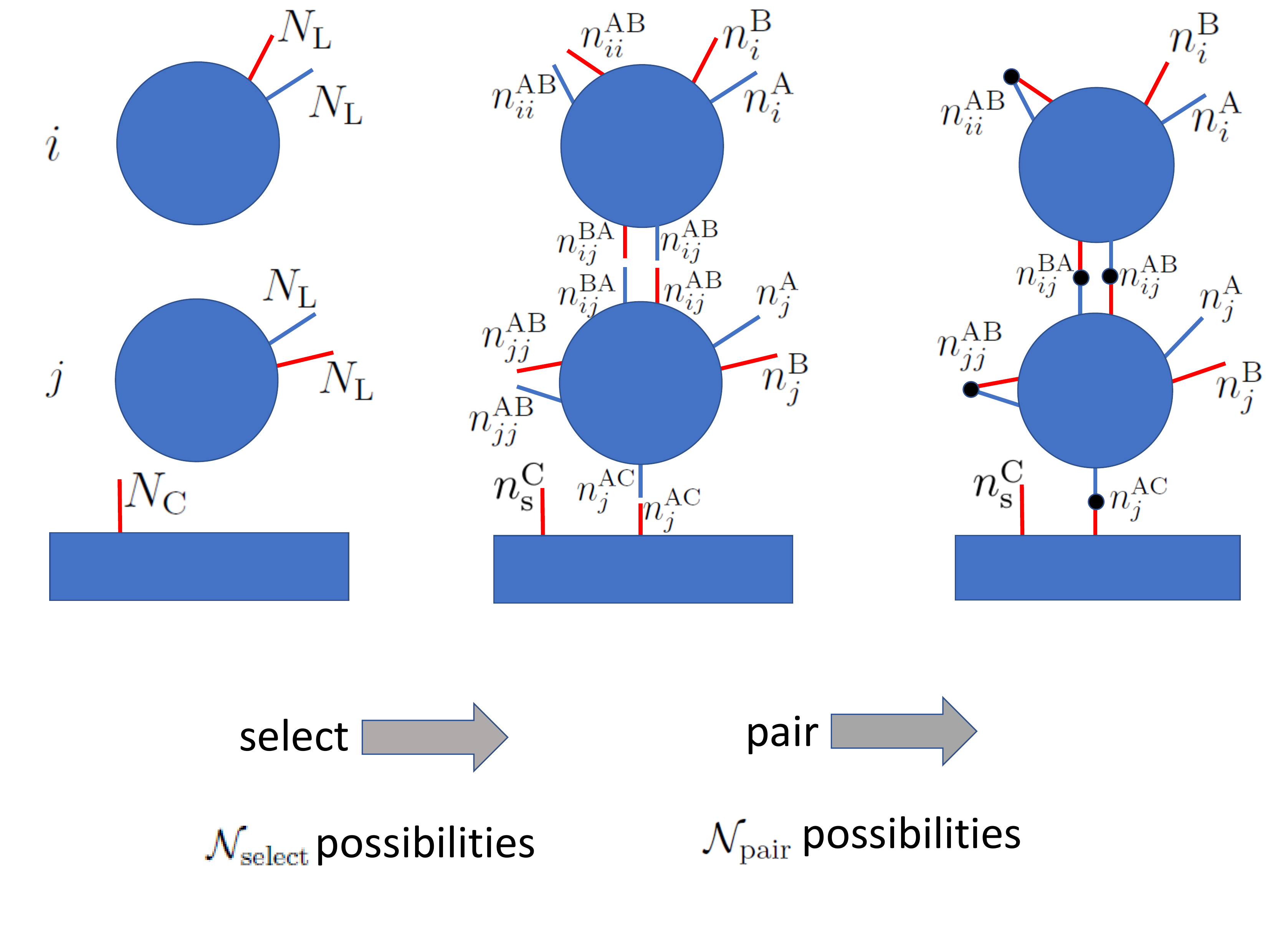}
\caption{\textbf{Counting the number of ways of forming a given set of linkages $\{n\}$.} ${\cal N}_\mthr{select}$ (see Main Eq.~6) is the number of ways of selecting the ligands/receptors to be used to form each type of linkage entering in $\{ n \}$. ${\cal N}_\mthr{pair}$ is the number of ways of binding the preselected ligands/receptors.}\label{Fig:Combinatoire}
\end{center}
\end{figure}

\begin{figure}[ht!]
\begin{center}
\includegraphics[width=6cm]{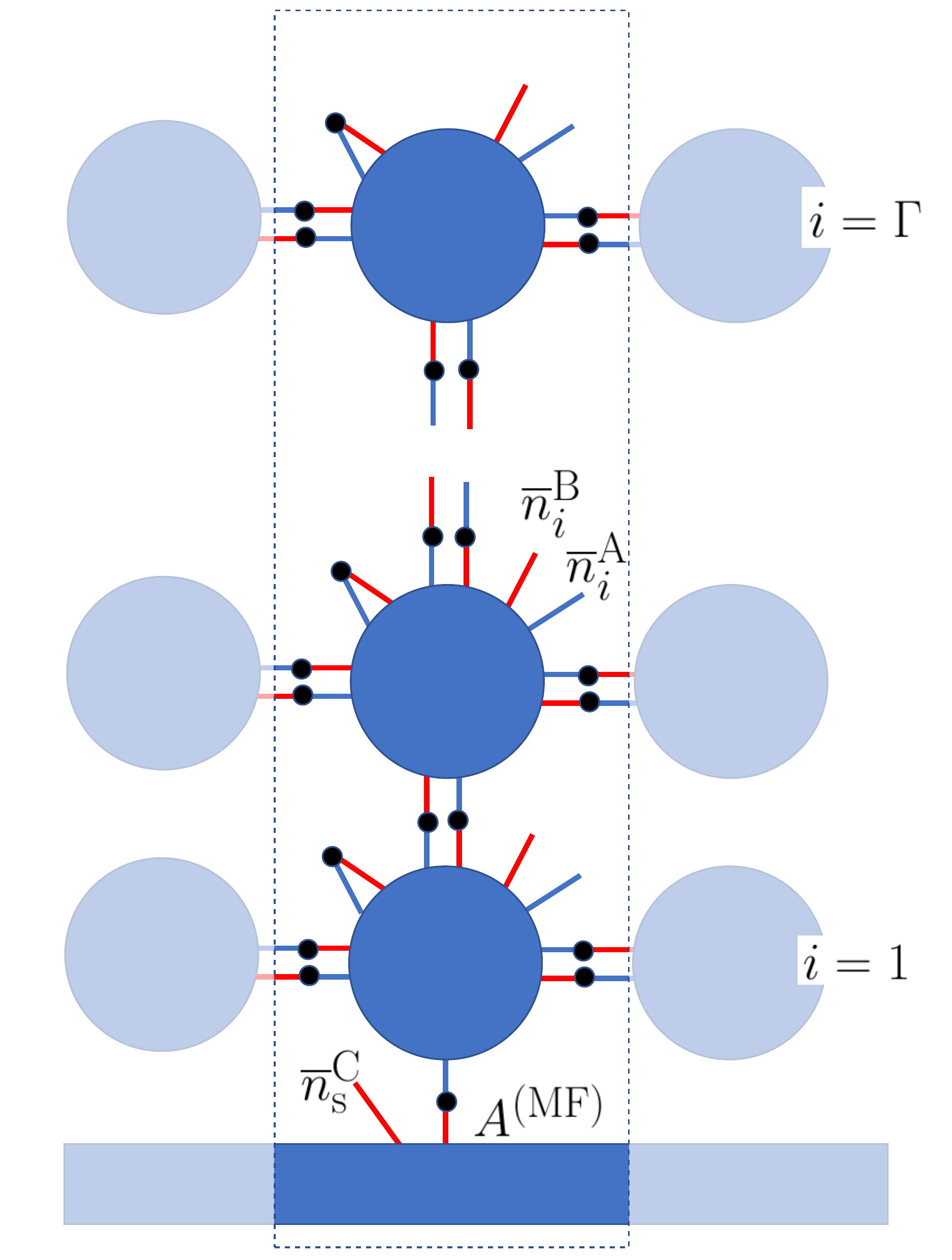}
\caption{\textbf{Definitions of the variables used in the mean field calculation.} We consider fcc (111) crystals in which each colloid is surrounded by six particles belonging to the same layers ($i$) and three particles belonging to the upper ($i+1$) and/or lower ($i-1$) layer (for simplicity the figure reports at most four neighbors). Particles belonging to a given layer $i$ feature the same type of linkages and carry the same number of free A and B binders, $\on^\mthr{A}_i$ and $\on^\mthr{B}_i$. The particle in the first layer faces a surface area $A^\mthr{(MF)}$ estimated using simulations. Accordingly, the total number of receptors $N^\mthr{(MF)}_\mthr{C}$ is defined by the receptor density $\sigma_\mthr{R}$. In Main Figure 5 we used $A^\mthr{(MF)}=10.31 \times 10.31 \cdot L^2$ while in Main Figure 6 $A^\mthr{(MF)}=11.0 \times 11.0 \cdot L^2$. Particle--particle and particle--surface distances have been fixed to  $11\cdot L$ and $5.7\cdot L$, respectively. 
}\label{Fig:MFT}
\end{center}
\end{figure}

\begin{figure}[ht!]
\begin{center}
\includegraphics[width=6cm]{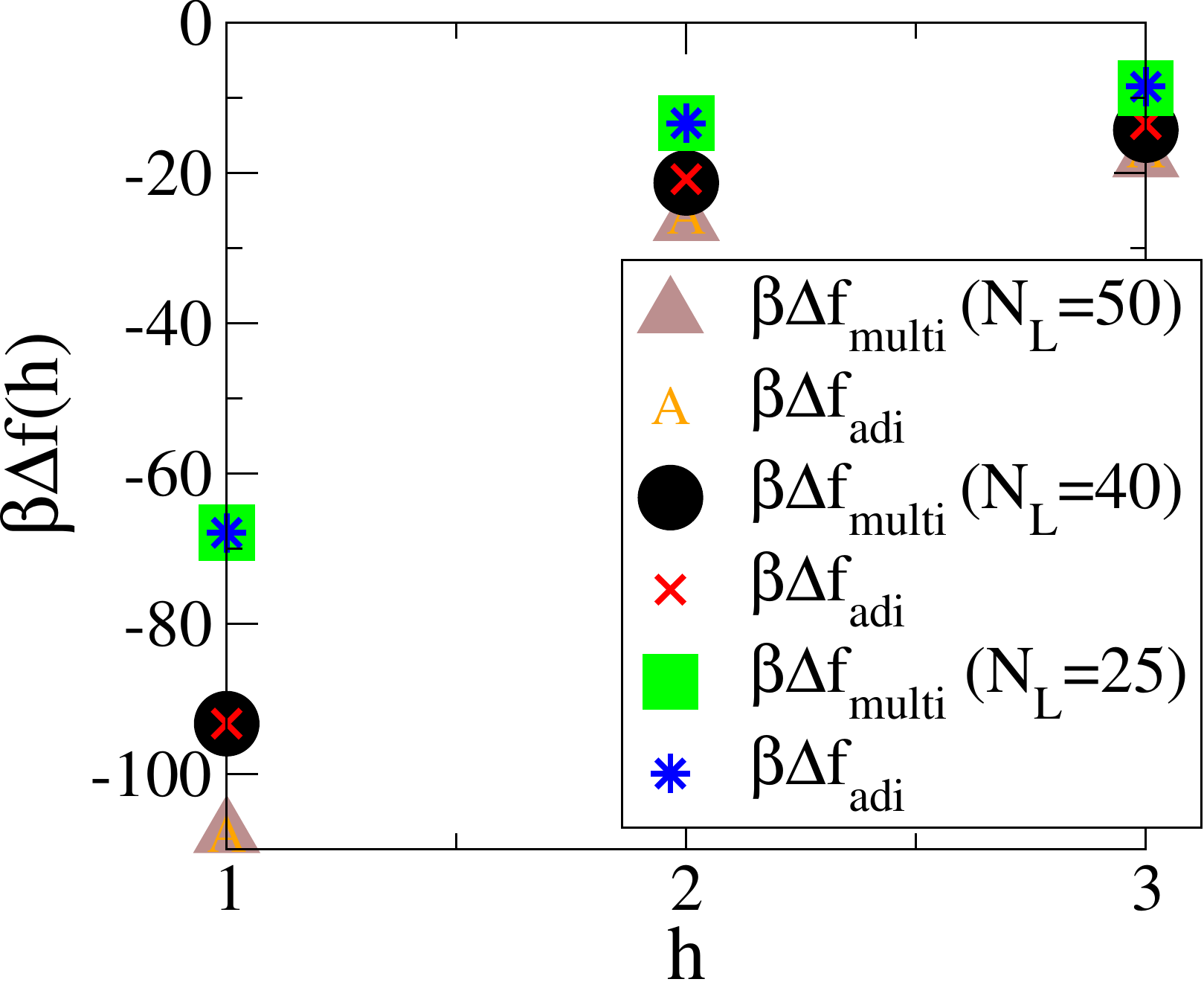}
\caption{\textbf{Analytic approximations of the multivalent free energy match the results of the full theory.} For the three different number of ligands ($N_\mathrm{L}$) considered in the present work, we compare the multivalent free energy as provided by the full theory ($\Delta f_\mathrm{multi}(h)$, see Eq.~\ref{eq:fmulti}) with the analytic predictions of a simplified theory ($\Delta f_\mathrm{adi}(h)$, see Eq.~\ref{eq:fadi}) proving their consistency. }\label{Fig:MFTb}
\end{center}
\end{figure}

\section*{Supporting Information}

\section*{S1: Calculation of the multivalent free energy of the system}

We calculate the partition function of the system at a given colloids' configuration $\{ {\bf r} \}$ and number of linkages $\{ n \}$ (see Main Eq.~1) leading to the multivalent free energy ${\cal F}_\mthr{multi}(\{ {\bf r} \},\{ n \})=-k_BT \log {\cal Z}(\{ {\bf r} \},\{ n \})$. 
We start by computing the number of ways of making a certain set of linkages, $\{n\}=\{n^\mthr{AB}_{ij}, \, n^\mthr{AB}_{ji}, \, n^\mthr{AB}_{ii}, \, n^\mthr{AC}_{i} \}$ with $i=1,\cdots N_\mthr{p}$ and $i<j$. 
First, we count the number of ways, ${\cal N}_\mthr{select}$, of selecting the binders (ligands and receptors) used to form a certain type of linkage (e.g., a bridge of type AB between particle $i$ and particle $j$, see left and central panel in Fig.~\ref{Fig:Combinatoire}). 
The number of ways of selecting $n^\mthr{AB}_{ij}$, $n^\mthr{AB}_{ii}$, and $n^\mthr{AC}_{i}$ A ligands on particle $i$ to be used to form, respectively, the bridges with particle $j$ ($j=1,\cdots, N_\mthr{p}$), the loops, and the surface--particle bridges is 
\begin{eqnarray}
{\cal N}^i_\mthr{select,A} &=& {N_\mthr{L} \choose n^\mthr{AB}_{ii}} {N_\mthr{L} -n^\mthr{AB}_{ii} \choose n^\mthr{AC}_{i}}
 \cdot\prod_{j\neq i} {N_\mthr{L} -n^\mthr{AB}_{ii}-n^\mthr{AC}_i- \sum_{p=1, p\neq i}^{j-1} n^\mthr{AB}_{ip} \choose n^\mthr{AB}_{ij}}
= {N_\mthr{L} ! \over n^\mthr{A}_i !  n^\mthr{AC}_i! n^\mthr{AB}_{ii}! \prod_{j\neq i} n^\mthr{AB}_{ij} ! } \, ,
\end{eqnarray}
where $n^\mthr{A}_i$ is the number of free A ligands
\begin{eqnarray}
n_i^\mathrm{A} &=&N_\mathrm{L}-n^{AB}_{ii}-n_i^\mathrm{AC}-\sum_{j\neq i} n_{ij}^\mathrm{AB}  \, .
\end{eqnarray}
 Similarly, the number of ways of selecting the B linkers on particle $i$ and the receptors are
\begin{eqnarray}
{\cal N}^i_\mthr{select,B} = {N_\mthr{L} ! \over n^\mthr{B}_i !  n^\mthr{AB}_{ii}! \prod_{j\neq i} n^\mthr{BA}_{ij} ! }  &\qquad \quad &   n_i^\mathrm{B} = N_\mathrm{L}-n^{AB}_{ii} -\sum_{j\neq i} n_{ij}^\mathrm{BA} ,  
\nonumber \\
{\cal N}_\mthr{select,C} = { N_\mthr{C} ! \over n^\mthr{C}_\mthr{s} !  \prod_{i=1}^{N_\mthr{p}} n^\mthr{AC}_i! }  & \qquad \quad &  n^\mathrm{C}_\mathrm{s} = N_\mathrm{C}-\sum_i n_i^\mathrm{AC} , 
\nonumber
\end{eqnarray}
where $n^\mthr{C}_\mthr{s}$ is the number of free receptors on the surface. Finally, ${\cal N}_\mthr{select}$ reads as 
\begin{eqnarray}
{\cal N}_\mthr{select} = {\cal N}_\mthr{select,C} \prod_{i=1}^{N_\mthr{p}} \left[ {\cal N}^i_\mthr{select,A} {\cal N}^i_\mthr{select,B} \right].
\end{eqnarray}
After selecting the ensemble of ligands/receptors to be used to form linkages $\{ n \}$, we calculate the number of ways, ${\cal N}_\mthr{pair}$, of reacting the pre--selected binders (see right panel of Fig.~\ref{Fig:Combinatoire}). Noticing that there are, for instance, $n^\mthr{AB}_{ij}!$ ways of binding $n^\mthr{AB}_{ij}$ A ligands on particle $i$ with $n^\mthr{AB}_{ij}$ B ligands on particle $j$, we obtain 
\begin{eqnarray}
{\cal N}_\mthr{pair} = \prod_i \left[ n^\mthr{AC}_i! n^\mthr{AB}_{ii}! \right] \prod_{i<j} \left[ n^\mthr{AB}_{ij} ! n^\mthr{BA}_{ij} ! \right] \, .
\end{eqnarray}
Finally, the combinatorial term associated to configurations with a given set of linkage $\{ n \}$ is 
\begin{eqnarray}
{\cal N}_\mthr{comb}& = &{\cal N}_\mthr{select} \times {\cal N}_\mthr{pair}
 =  {N_\mthr{C}! \over n^\mthr{C}_\mthr{s}! } \prod_i \left[{N_\mthr{L} ! \over n^\mthr{A}_i ! } {N_\mthr{L} ! \over n^\mthr{B}_i ! } {1 \over n^\mthr{AB}_{ii}! } {1\over n^\mthr{C}_\mthr{s} ! } \right] 
\prod_{i<j} \left[ {1 \over n^\mthr{AB}_{ij} !}{1 \over n^\mthr{BA}_{ij} !} \right]
\nonumber
\end{eqnarray}
To derive the final expression of ${\cal Z}$, we multiply ${\cal N}_\mathrm{comb}$ by the Boltzmann factor accounting for the hybridization free energy of binding a set $\{ n \}$ of linkages (see Main text for the definitions)
\begin{eqnarray}
  {\cal Z} &=& 
\left[ \prod_{i=1}^{N_\mathrm{p}}  \frac{N_\mathrm{L}! N_\mathrm{L}! e^{-\beta n^{\mathrm{AB}}_{ii} \Delta G_{ii}(\{{\bf{r}}\})}  }{ n^\mathrm{A}_i! n^\mathrm{B}_i! n^{\mathrm{AB}}_{ii}!} \right] \label{Equ:Z}
\left[ \prod_{j<q}   \frac{ e^{-\beta (n^{\mathrm{AB}}_{jq} + n^{\mathrm{BA}}_{jq} ) \Delta G_{jq}(\{{\bf{r}} \})  } }{ n^{\mathrm{AB}}_{jq}! n^{\mathrm{BA}}_{jq}! } \right]  
 \left[
\frac{N_\mathrm{C}!}{ n_\mathrm{s}^\mathrm{C}! }\prod_{i=1}^{N_\mthr{p}}  \frac{ e^{-\beta n^{\mathrm{AC}}_i  \Delta G^{s}_i (\{{\bf{r}} \})  } }{ n^{\mathrm{AC}}_i ! } 
\right] 
 \cdot {\cal Z}_{T=\infty}(\{{\bf{r}}\}) \, .
\label{Eq:Z}
\end{eqnarray} 
${\cal Z}_{T=\infty}$ denotes the partition function of the system when no linkages are formed (as found at high temperature). ${\cal Z}_{T=\infty}$ is related to entropic, repulsive forces due to the reduction of the configurational space of the DNA linkers compressed by two approaching surfaces.
Neglecting excluded volume interactions between linkers, as often done when modeling DNA mediated interactions, and defining by $\Omega_0$ and $\Omega_i (\{ {\bf r} \})$ the volume of the configurational space available, respectively, to a single ligand tethered to particle $i$ at infinite dilution and at finite density (similar definitions follow for the configurational space of receptors tethered to the surface, $\Omega^\mathrm{s}_\mathrm{0}$ and $\Omega^\mathrm{s}(\{ {\bf r} \})$) we find
\begin{eqnarray}
Z_{T=\infty}(\{{\bf{r}}\}) &=& e^{-\beta F_{T=\infty} (\{{\bf{r}}\})}= \left( \frac{\Omega^\mathrm{s} (\{ {\bf{r}}\}) }{ \Omega^\mathrm{s}_\mathrm{0} } \right)^{N_\mathrm{C}} \cdot
  \prod_{i=1}^{N_\mathrm{p}} \left( \frac{\Omega_i (\{ {\bf{r}}\}) }{ \Omega_0 }\right)^{N_\mathrm{L}+N_\mathrm{L}}  
 e^{-\beta V(\{{\bf{r}}\})} \,  .
\label{Eq:ZInf}
\end{eqnarray}
In the previous expression, $V(\{ {\bf r} \})$ accounts for additional (e.g.\ electrostatic) interactions between colloids. We use $V(\{ {\bf r} \}))$ to regularize hard--core repulsions with negligible effects on the final results. Below (see Sec.~S1.A), we report the expression of $V(\{ {\bf r} \})$ that has been used in this work.
\\
In this work we consider reactive sequences tethered to particles' surfaces through short, thin rods of double-stranded DNA of length $L$. When $L$ is much smaller than the radius of the particles, the reactive sequences of unbound linkers are uniformly distributed within the layer of thickness $L$ surrounding the tethering surfaces. We then have $\Omega_0 = 4 \pi R^2 L $, $\Omega^\mathrm{s}_0={\cal A} L$ (where ${\cal A}$ is the area of the surface). Similarly, the configurational volumes defining the hybridization free energies (see Main Eqs.~4) read as 
\begin{eqnarray}
\Omega_i(\{ {\bf r} \}) = \Omega_0 - e^\mathrm{s}_i(r_{i,z}) - \sum_{j\in v(i)} e_{ij}(|{\bf r}_i - {\bf r}_j|)  
& \qquad &
\Omega^\mathrm{s}(\{ {\bf r} \}) = \Omega^\mathrm{s}_0 - \sum_{i \in v_\mathrm{s}} k^\mathrm{s}_i(r_{i,z}) 
\label{Eq:Omega}
\end{eqnarray}
where $e_{ij}$ and $e^\mathrm{s}_i$ are, respectively, the volume excluded to the reactive sequence of a linker tethered to particle $i$ by the presence of particle $j$ and the surface (see Main Fig.~2). $k^\mathrm{s}_i$ is the volume excluded to a reactive sequence tethered to the surface by the presence of particle $i$ (see Main Fig.~2). $v(i)$ and $v_\mathrm{s}$ are the lists of particles in direct contact with ligands on particle $i$ and receptors on the surface. Similarly, the configurational space of bound sequences ($\Omega_{ij}$ and $\Omega^\mthr{s}_i$) is the volume of the overlapping regions spanned by the reacting sequences before binding (see Main Fig.~2). We report the explicit expression of the terms appearing in Main Eqs.~4 and Eqs.~\ref{Eq:Omega} in Sec.~S1.B.
\\
At given colloid positions, $\{ {\bf r} \}$, the most likely number of linkages featured by the system, $\{ \overline n \}$, are calculated by maximizing the multivalent free energy 
\begin{eqnarray}
\frac{\partial }{\partial \{ n\} } {\cal F}_\mathrm{multi} (\{n\})|_{\{n\} = \{ \on \}} = 0 \, .
\end{eqnarray}
The previous set of equations, along with the definition of ${\cal F}_\mathrm{multi}$, Eq.\ \ref{Equ:Z}, lead to the chemical equilibrium equations reported in Main Eqs.~3. Main Eqs.~3 become equivalent to the following set of equations for the number of unbound linkers 
\begin{eqnarray}
\on^\mathrm{A}_i &=& \frac{N_\mathrm{L} }{ 1 + \on^\mathrm{C}_\mthr{s} e^{-\beta \Delta G^\mthr{s}_{i}(\{{\bf r}\})} + \sum_{j\in v(i)} \on^\mathrm{B}_j e^{-\beta \Delta G_{ij}(\{{\bf r}\})} }
\nonumber \\
\on^\mathrm{B}_i &=& \frac{N_\mathrm{L} }{ 1 + \sum_{j\in v(i)} \on^\mathrm{A}_j e^{-\beta \Delta G_{ij}(\{{\bf r}\})}  }
\nonumber \\
\on^\mathrm{C}_\mathrm{s}
&=& \frac{N_\mathrm{C} }{ 1+ \sum_{j\in v_\mthr{s} } \on^\mathrm{A}_j e^{-\beta \Delta G^\mathrm{s}_j (\{{\bf r}\})} } \, ,
\label{Eq:ChemESI}
\end{eqnarray}
that are used to implement self-consistent calculations in our simulations (see Sec.~\ref{Sec:Sim}). When written in term of the stationary number of linkages, the multivalent free energy simplifies into the expression reported in Main Eq.~5.

\subsection*{S1.A: Modeling hard--core repulsion}

We use smooth pair potentials to regularize particle--particle and particle--surface hard--core repulsions ($V(\{ {\bf r}\})$ in the Methods section of the main text). Following previous investigations, the repulsion between colloids is modeled using
\begin{eqnarray}
V_\mthr{pp}(r_{ij}) &=& \left\{
\begin{array}{ll}
500 \log \left( 1- {e_{ij}(r_{ij},0.75 \cdot L) \over 4 \pi R^2 \cdot (0.75 \cdot L)
} \right) &  r_{ij} < 2 \cdot R+0.75 \cdot L 
\\
0 &  r_{ij} \geq 2 \cdot R+0.75 \cdot L 
\end{array}
\right.
\end{eqnarray}
where $r_{ij}$ is the distance between particle $i$ and $j$, and $e_{ij}$ is defined below (Eq.~\ref{Eq:eij}). Similarly, if $r_{i,\mthr{z}}$ is the distance of particle $i$ from the surface, the surface-particle repulsion is modeled using
\begin{eqnarray}
V_\mthr{ps}(r_{i,\mthr{z}}) &=&  
\left\{
\begin{array}{ll}
500 \log \left( 1- {e^\mthr{s}_i (r_{i,\mthr{z}},0.75 \cdot L) \over 4 \pi R^2 \cdot (0.75 \cdot L)
} \right) &  r_{i,\mthr{z}} <  R+0.75 \cdot L 
\\
0 &  r_{i,\mthr{z}} \geq   R+0.75 \cdot L 
\end{array}
\right.
\, ,
\end{eqnarray}
where $e^\mthr{s}_i$ is defined in Eq.~\ref{Eq:esi}. Finally 
\begin{eqnarray}
V(\{ {\bf r } \}) = \sum_{i<j} V_\mthr{pp}(r_{ij})  + \sum_i V_\mthr{ps}(r_{i,\mthr{z}}) \, . 
\end{eqnarray}
Such smooth regularizations allow using larger integration steps $\Delta t$ (see the Methods section of the main text) with negligible effects on the results of the manuscript. The latter claim follows from the fact that the typical surface--to--surface distance is comparable with $L$ while $V_\mthr{pp}(r_{ij})>0$ and $V_\mthr{ps}(r_{i\mthr{z}})>0$ only when $r<0.75 \cdot L$. $V_\mthr{pp}$ and $V_\mthr{ps}$ are effective potentials resulting from coating particles with inert strands (not carrying sticky ends) of length $0.75 \cdot L$. This observation justifies the particular choice of $V$ given the fact that, in experiments, inert binders are often used to screen non-selective attractions (e.g., van der Waals forces).

\subsection*{S1.B: Configurational terms}

Below we report the analytic expressions of the overlapping volumes defined in Main Fig.~2$b$ and used to calculate the hybridization free energies and the $on$ rates (see the Methods section in the main text) 
\begin{eqnarray}
\Omega_{ij} (r_{ij},L) &=& v_\mthr{par,par}(r_{ij},R+L,R+L) - 2 v_\mthr{par,par}(r_{ij},R,R)
\label{eq:omegaij}
\\
\Omega^\mthr{s}_i (r_{i,\mthr{z}},L) &=& v_\mthr{par,surf}(r_{i,\mthr{z}}-L,R+L) - v_\mthr{par,surf}(r_{i,\mthr{z}},R+L)  - v_\mthr{par,surf}(r_{i,\mthr{z}}-L,R) 
\\
e_{ij} (r_{ij},L) &=& v_\mthr{par,par}(r_{ij},R+L,R) \label{Eq:eij}
\\
e^\mthr{s}_i (r_{i,\mthr{z}},L) &=& v_\mthr{par,surf}(r_{i,\mthr{z}},R+L)  \label{Eq:esi}
\\
k^\mthr{s}_i (r_{i,\mthr{z}},L) &=& v_\mthr{par,surf}(r_{i,\mthr{z}}-L,R)
\end{eqnarray}
where $v_\mthr{par,par}(r,R_1,R_2)$ and $v_\mthr{par,surf}(r,R)$ are, respectively, the overlapping volume between two spheres of radius $R_1$ and $R_2$ placed at distance $r$ and the volume of a spherical cap of radius $R$ with base placed at distance $r$ from the center of the sphere:
\begin{eqnarray}
v_\mthr{par,par}(r,R_1,R_2) &=& {\pi \over 12 r } \left( R_1+R_2-r \right)^2 \left( r^2 + 2 r R_1 + 2r R_2 -3 R_1^2 - 3 R_2^2 + 6R_1 R_2 \right)
\\
v_\mthr{par,surf}(r,R) &=&  {\pi \over 3 } \left( R-r\right)^2 \left(2R+r\right) 
\end{eqnarray}

\section*{S2: Mean Field Theory}

We detail the calculation of $f_\mthr{multi}$ (see Main Eqs.~9) used to predict the probability of self-assembling crystals with $\Gamma$ layers (see $P(\Gamma)$ defined in Main Eq.~10). $f_\mathrm{multi}(\Gamma)$ is the free energy of an fcc (111) crystallite comprising $\Gamma$ layers as compared to a reference state in which particles are isolated and only feature loops, normalized by the number of particles in direct contact with the functionalized surface.
 We first calculate the number of free binders and free receptors on the particles belonging to layer $i$ along with the number of free receptors ($\on^\mthr{A}_i$, $\on^\mthr{B}_i$, and $\on^\mthr{C}_\mthr{s}$ in Fig.~\ref{Fig:MFT}).
 We consider particles distributed on an fcc~(111) crystal with fixed particle--particle and particle--surface distance. Therefore, the free energy of making inter--particle bridges ($\Delta G^\mthr{b}$) is constant and is calculated using Main Eq.~4. Similarly, $\Delta G^\mthr{l}$ and $\Delta G^\mthr{b,s}$ are the hybridization free energies of forming loops and particle--surface bridges. We calculate $\Delta G^\mthr{b,s}$ using a surface area $A^\mthr{(MF)}$ corresponding to the averaged surface {\it per} surface--bound particle as sampled in a representative simulation (notice that $A^\mthr{(MF)}$ affects $\Delta G^\mthr{b,s}$ through $\Omega^\mthr{s}$, see Main Eq.~4). Accordingly, we fix the number of receptors $N^\mthr{(MF)}_\mthr{C}$ to $N^\mthr{(MF)}_\mthr{C}=\sigma_R \cdot A^\mthr{(MF)}$, where $\sigma_R$ is the receptor density.
\\
Using Eqs.~\ref{Eq:ChemESI} we write the total number of free binders on the surface and on the particles belonging to the first layer as 
\begin{eqnarray}
\on_\mathrm{s}^\mathrm{C} &=& { N^\mthr{(MF)}_\mathrm{C} \over 1+ \on^\mthr{A}_i  \exp[-\beta \Delta G^\mathrm{b,s}] } \nonumber
\\
\on_1^\mathrm{A} &=& {N_\mathrm{L} \over 1+ \on_1^\mathrm{B} (6\cdot \exp[-\beta \Delta G^\mthr{b}] + \exp[-\beta \Delta G^\mthr{l}]) + 3\cdot  \on_2^\mathrm{B}  \exp[-\beta \Delta G^\mthr{b}]  + \on_\mathrm{s}^\mathrm{C}  \exp[-\beta \Delta G^\mathrm{b,s}]}
\nonumber \\
\on_1^\mathrm{B} &=& {N_\mathrm{L} \over 1+   \on_1^\mathrm{A} (6\cdot \exp[-\beta \Delta G^\mthr{b}]  + \exp[-\beta \Delta G^\mthr{l}]) + 3 \cdot \on_2^\mathrm{A} \exp[-\beta \Delta G^\mthr{b}] }
\nonumber
\end{eqnarray}
where in the expression of $\on^\mthr{A}_1$ and $\on^\mthr{B}_1$ we set $\on^\mthr{A}_2=\on^\mthr{B}_2=0$ when calculating the free energy of single--layer crystals ($\Gamma=1$ in Fig.~\ref{Fig:MFT}). For particles in the intermediate layers $1< i < \Gamma$ we have 
\begin{eqnarray}
\on^\mathrm{A}_i &=& {N_\mathrm{L} \over 1+ \on_i^\mathrm{B} (6\cdot \exp[-\beta \Delta G^\mthr{b}]  + \exp[-\beta \Delta G^\mthr{l}]) +3 \cdot  (\on_{i-1}^\mathrm{B} + \on_{i+1}^\mathrm{B}) \exp[-\beta \Delta G^\mthr{b}] }
\nonumber \\
\on_i^\mathrm{B} &=& {N_\mathrm{L} \over 1+ \on_i^\mathrm{A} (6\cdot \exp[-\beta \Delta G^\mthr{b}]  + \exp[-\beta \Delta G^\mthr{l}]) + 3\cdot (\on^\mathrm{A}_{i-1} +\on^\mathrm{A}_{i+1})  \exp[-\beta \Delta G^\mthr{b}] }
\nonumber
\end{eqnarray}
while for the last layer (if $\Gamma>1$)
\begin{eqnarray}
\on^\mathrm{A}_\Gamma &=& {N_\mathrm{L} \over 1+ \on_\Gamma^\mathrm{B} (6\cdot \exp[-\beta \Delta G^\mthr{b}]  + \exp[-\beta \Delta G^\mthr{l}]) + 3\cdot \on_{\Gamma-1}^\mathrm{B} \exp[-\beta \Delta G^\mthr{b}]  }
\nonumber
\\
\on^\mathrm{B}_\Gamma &=& {N_\mathrm{L} \over 1+ \on^\mathrm{A}_\mathrm{\Gamma} (6 \cdot \exp[-\beta \Delta G^\mthr{b}]  + \exp[-\beta \Delta G^\mthr{l}]) + 3 \cdot \on^\mathrm{A}_{\Gamma-1} \exp[-\beta \Delta G^\mthr{b}] }
\nonumber
\end{eqnarray}
We define by $\on^\mthr{A}_0$ and $\on^\mthr{B}_0$ ($\on^\mthr{A}_0=\on^\mthr{B}_0$) the number of free binders of particles isolated in bulk (therefore featuring $N_\mthr{L}-\on^\mthr{A}_0$ loops). $\on^\mthr{A}_0$ or $\on^\mthr{B}_0$ is calculated by setting $\Delta G^\mthr{b}=+\infty$ and $\Delta G^\mathrm{b,s}=+\infty$ in one of the previous equations.
\\
The free energy of $\Gamma$ isolated particles in bulk reads as (see Main Eq.~5)
\begin{eqnarray}
\beta F_0 (\Gamma)= \Gamma \left[ 
N_L \log  {\on^\mthr{A}_0 \over N_\mthr{L}}  + N_L \log {\on^\mthr{B}_0 \over N_\mthr{L}} + (N_\mthr{L} - \on^\mthr{A}_0)
\right] \, , 
\end{eqnarray}
where $N_\mthr{L} - \on^\mthr{A}_0$ is the number of loops featured by each particle. 
On the other hand, the free energy {\it per} surface--bound particle of crystals with $\Gamma$ layers (see Fig.~\ref{Fig:MFT}) is given by (see Main Eq.~1)
\begin{eqnarray}
\beta F(\Gamma) &=&  N_\mathrm{C} \log {\on_\mathrm{s}^\mathrm{C} \over N_\mathrm{C} } + {N_\mathrm{C}-\on_\mathrm{s}^\mathrm{C}\over 2 } + \sum_{i=1}^\Gamma \left[N_\mathrm{L}\log {\on_i^\mathrm{A} \over N_\mathrm{L}} +  N_\mathrm{L}\log {\on_i^\mathrm{B} \over N_\mathrm{L}} + {(N_\mathrm{A}+N_\mathrm{B}) - \on_i^\mathrm{A}-\on^\mathrm{B}_i\over 2}\right]\, .
\nonumber \\
\end{eqnarray}
Notice that in the configuration of Fig.~\ref{Fig:MFT} particles are sufficiently distanced that $F_{T=\infty}=0$. Finally,  $f_\mthr{multi}(\Gamma)$ (Main Eq.~9) used in the definition of $P$ (Main Eq.~10) and $\Delta f_\mathrm{multi} (\Gamma)$ (see Main Fig.~5a-c) are given by 
\begin{eqnarray}
f_\mthr{multi}(\Gamma) = F(\Gamma) - F_0 (\Gamma) & \qquad &   \Delta f_\mthr{multi}(\Gamma)= f_\mthr{multi}(\Gamma) -  f_\mthr{multi}(\Gamma-1) \, . 
\label{eq:fmulti}
\end{eqnarray}
To calculate the equilibrium layer distribution $P(\Gamma)$, we also need to estimate the entropic loss of caging colloids from the gas phase into a site of the crystal. We employ a cell model in which we assign a configurational space volume equal to $v_0$ to each particle in the solid phase. Following Main Refs.~[42,43] we use $v_0=(L/2)^3$, where we identify $L$ with the interaction range of a square well potential. We study the sensitivity of our results to $v_0$ in Main Fig.~8. 
The probability of assembling $\Gamma$ layers from a diluted colloidal suspension at density $\rho_\mathrm{id}$ reads as $P(\Gamma) ={1 / Z_\mthr{MFT}}\cdot \exp[-\beta f_\mathrm{multi}(\Gamma) ](\rho_\mathrm{id} v_0)^\Gamma$, where $\rho_\mathrm{id} \sim \exp[\beta \mu]$, $\mu$ is the chemical potential of the particles, and $\rho_\mathrm{id}$ is small enough to justify an ideal representation of the gas phase. 
\\
$Z_\mthr{MFT}$ is a normalization factor that is well defined (i.e.\ $Z_\mthr{MFT}=\sum_{\Gamma=0}^\infty \exp[-\beta f_\mathrm{multi}(\Gamma) ](\rho_\mathrm{id} v_0)^\Gamma  < \infty$) conditional on the gas phase being stable in bulk. To extract the phase boundary in bulk,  we notice that for $\Gamma \to\infty$ surface effects are negligible and $f_\mathrm{multi}(\Gamma)$ reads as $\Delta f (\infty) \cdot \Gamma$, where $ \Delta f_\mthr{multi} (\infty)$, $ \Delta f_\mthr{multi} (\infty) = \lim_{\Gamma \to \infty} f_\mthr{multi} (\Gamma) -f_\mthr{multi} (\Gamma-1)$, is the multivalent free energy {\it per} particle in an fcc crystal as compared to the gas phase. The gas--solid boundary in bulk (see Main Figs.~3$b$, 3$c$, and 4) is given by the relation $\exp[-\beta \Delta f_\mthr{multi}(\infty)] \rho_\mathrm{id} v_0 =1$ or $\beta \Delta f_\mthr{multi} (\infty) = \log (\rho_\mathrm{id} v_0) $ that, in the diluted limit $\rho_\mathrm{id}\to 0$, matches existing cell models Main Ref.~[43].

\subsection*{S2.A: Analytic predictions}
In this section we extract compact analytic expressions allowing to estimate $\Delta f_\mathrm{multi}$ (see Eq.~\ref{eq:fmulti}). We consider the low temperature regime in which $ \Delta G_0 \to -\infty$. Using Main Eq.~5, we calculate the statistical weight of, respectively, particle--surface and particle--particle bridges divided by the statistical weight of intra--particle loops as
\begin{eqnarray}
\chi^\mathrm{b,s} &=& { \exp[-\beta \Delta G^\mathrm{b,s}] \over \exp[-\beta \Delta G^\mathrm{l}]} = {\Omega^\mthr{s}_i \over \Omega^\mthr{s}_0 } 
\nonumber \\
\chi^\mathrm{b} &=& { \exp[-\beta \Delta G^\mathrm{b}] \over \exp[-\beta \Delta G^\mathrm{l}]} ={\Omega_{ij} \over \Omega_0 } \, \,  .
\end{eqnarray}
In the previous expressions, we considered that for the colloidal arrangement used in the MFT (see Fig.~\ref{Fig:MFT}) the configurational space of free binders is not excluded by any colloid or the surface ($\Omega_i=\Omega_0$ and $ \Omega^\mthr{s}= \Omega^\mthr{s}_0$). Note that $\Omega_{ij}$ and $\Omega^\mthr{s}_i$ are not a function of the specific particle $i$ given that particle--particle and particle--surface distances are kept constant (see Eq.~\ref{eq:omegaij}).  
Once the first layer of particles is formed, each particle will present a number of free ligands equal to 
\begin{eqnarray}
M_1 &=& { \sqrt{(\chi^\mathrm{b,s})^2(N^\mthr{(MF)}_\mthr{C}-N_\mthr{L})^2+4\chi^\mathrm{b,s} N^\mthr{(MF)}_\mthr{C} N_\mthr{L} (1+ 6 \chi^\mathrm{b})} - \chi^\mathrm{b,s} (N^\mthr{(MF)}_\mthr{C}+N_\mthr{L}) \over 2(1-\chi^\mathrm{b,s}+6 \chi^\mathrm{b})} \, \, .
\nonumber \\
\end{eqnarray}
We now assume that when particles are added to the second layer, the number of particle--surface bridges as well as the number of lateral bridges between particles in the first layer do not change. Such approximation allows calculating $\Delta f_\mthr{multi}(2)$ only using $M_1$. Similarly, by calculating the number of free linkers featured by particles in the second layer ($M_2$, which is only function of $M_1$), we can re-iterate the calculation of $\Delta f_\mthr{multi}(3)$ for particles in the third layer. In general, the free-energy gain of adding layer $i$ when particles in layer $i-1$ express $M_{i-1}$ free linkers is  
\begin{eqnarray}
\Delta f_\mathrm{adi} (i) &=& {3 \chi^\mathrm{b} \left[ M_{i-1}- 3 (N_\mthr{L} + 4 \chi^\mathrm{b} N_\mthr{L}) \right] - \Lambda \over 4 (1+3  \chi^\mathrm{b}) (1+6  \chi^\mathrm{b})} + N_\mthr{L} \log {1 \over 3 \chi^\mathrm{b} (M_{i-1}- N_\mthr{L}) +\Lambda}
\nonumber \\
&& + N_\mthr{L} \log { \Lambda -3  \chi^\mathrm{b} (M_{i-1}+N_\mthr{L}) \over 1 + 3 \chi^\mathrm{b} } 
\label{eq:fadi}
\end{eqnarray}
where 
\begin{eqnarray}
\Lambda &=& \sqrt{ 3 \chi^\mathrm{b} } \sqrt{4 M_{i-1} N_\mthr{L} + 3 \chi^\mathrm{b} \left[ 
(M_{i-1})^2+6 M_{i-1} N_\mthr{L} + N_\mthr{L}^2
\right] } \, .
\end{eqnarray}
The number of free ligands {\em per} particle expressed by layer $i$ before attaching layer $i+1$ reads as
\begin{eqnarray}
M_i &=& {\Lambda -3 \chi^\mathrm{b} M_i -3 \chi^\mathrm{b} N_L \over 2 (1+ 3 \chi^\mathrm{b}) } \, \, .
\end{eqnarray}
Fig.~\ref{Fig:MFTb} shows that the results of the analytic theory match the MFT predictions. 

\section*{S3: Simulation details}\label{Sec:Sim} 
We first calculate the force acting on particle $i$, ${\bf f}_i$ in Main Eq.~6 and 7. When using the implicit (IMP) scheme (see Main Sec.~2.2), the numbers of possible linkages are fixed to their most likely values, $\{ n\} = \{ \on\}$ (see Main Eqs.~12, 13). The force acting on particle $i$ then reads as 
\begin{eqnarray}
\beta {\bf f}_i &=& -\beta {\bf \nabla}_{{\bf r}_i} {\cal F}_\mthr{multi} (\{ \on\} , \{ {\bf{r}} \}) 
=-{\bf \nabla}_{{\bf r}_i} \log {\cal Z} (\{ \on\} , \{ {\bf{r}} \})
\nonumber \\
&=& - {\partial \over \partial \{ n \} } \beta {\cal F}_\mthr{multi}(\{n\} , \{ {\bf{r}} \})|_{\{n\}=\{\on\}} {\bf \nabla}_{{\bf r}_i} \{ \on \} - {\partial \over \partial \{ \Delta G \} } \beta {\cal F}_\mthr{multi}(\{n\} , \{ {\bf{r}} \}) {\bf \nabla}_{{\bf r}_i} \{ \Delta G \}|_{\{n\}=\{\on\}} 
\nonumber \\
&& - \beta {\partial \over \partial {\bf r}_i} {\cal F}_\mthr{multi} (\{n\} , \{ {\bf{r}} \}) |_{\{n\}=\{\on\}}
\nonumber \\
&=& - {\partial \over \partial \{ \Delta G \} } \beta {\cal F}_\mthr{multi}(\{n\} , \{ {\bf{r}} \}) {\bf \nabla}_{{\bf r}_i} \{ \Delta G \}|_{\{n\}=\{\on\}} - \beta {\partial \over \partial {\bf r}_i} F_{T=\infty} ( \{ {\bf{r}} \}) 
\label{Equ:force_1}
\end{eqnarray}
where the second equality follows from the saddle point equations (Main Eq.~2) and the fact that the only direct dependency of ${\cal F}_\mthr{multi}$ on $\{ {\bf r} \}$ is due to $F_{T=\infty}$ (see Eqs.~\ref{Eq:Z} and \ref{Eq:ZInf}). In particular, we find
\begin{eqnarray}
 \beta {\bf f}_i &=& -\sum_{j=1}^{N_\mathrm{p}} \left[  \on^{\mathrm{AB}}_{jj} {\bf \nabla}_{{\bf r}_i}  \beta \Delta G_{jj}(\{ {\bf{r}} \}) + \on^{\mathrm{AC}}_j {\bf \nabla}_{{\bf r}_i}  \beta \Delta G^s_{j}(\{ {\bf{r}} \}) - (N_\mathrm{L}+N_\mathrm{L}) { {\bf \nabla}_{{\bf r}_i} \Omega_j(\{ {\bf{r}} \}) \over \Omega_j(\{ {\bf{r}} \}) } \right] 
\nonumber \\
&& -\sum_{1\leq j<q\leq N_\mathrm{p}} (  \on^{\mathrm{AB}}_{jq} +  \on^{\mathrm{BA}}_{jq} ) {\bf \nabla}_{{\bf r}_i}  \beta \Delta G_{jq}(\{ {\bf{r}} \})  + 
N_\mathrm{C} { {\bf \nabla}_{{\bf r}_i} \Omega^\mathrm{s} (\{ {\bf{r}} \}) \over \Omega^\mathrm{s}(\{ {\bf{r}} \}) }
- {\bf \nabla}_{{\bf r}_i} \beta V(\{{\bf{r}}\}) \, .
\label{Equ:force_2}
\end{eqnarray}
By using the definitions of $\Delta G$ we find 
\begin{eqnarray}
\beta {\bf \nabla}_{{\bf r}_i} \Delta G_{jj}(\{ {\bf{r}} \}) &=&  { {\bf \nabla}_{{\bf r}_i} \Omega_{j}(\{ {\bf{r}} \}) \over \Omega_{j}(\{ {\bf{r}} \}) }
 \\
\beta {\bf \nabla}_{{\bf r}_i} \Delta G_{jq}(\{ {\bf{r}} \}) &=&  { {\bf \nabla}_{{\bf r}_i} \Omega_{j}(\{ {\bf{r}} \}) \over \Omega_{j}(\{ {\bf{r}} \}) } + { {\bf \nabla}_{{\bf r}_i} \Omega_{q}(\{ {\bf{r}} \}) \over \Omega_{q}(\{ {\bf{r}} \}) } - { {\bf \nabla}_{{\bf r}_i} \Omega_{jq}(\{ {\bf{r}} \}) \over \Omega_{jq}(\{ {\bf{r}} \}) }
 \\
\beta {\bf \nabla}_{{\bf r}_i} \Delta G^\mathrm{s}_{j}(\{ {\bf{r}} \}) &=&  { {\bf \nabla}_{{\bf r}_i} \Omega_{j}(\{ {\bf{r}} \}) \over \Omega_{j}(\{ {\bf{r}} \}) } + { {\bf \nabla}_{{\bf r}_i} \Omega^\mathrm{s}(\{ {\bf{r}} \}) \over \Omega^\mathrm{s}(\{ {\bf{r}} \}) } - { {\bf \nabla}_{{\bf r}_i} \Omega^\mathrm{s}_{j}(\{ {\bf{r}} \}) \over \Omega^\mathrm{s}_{j}(\{ {\bf{r}} \}) }
\\
\beta {\bf \nabla}_{{\bf r}_i} \Delta G_\mathrm{s}(\{ {\bf{r}} \}) &=& { {\bf \nabla}_{{\bf r}_i}\Omega^\mathrm{s}(\{ {\bf{r}} \}) \over \Omega^\mathrm{s}(\{ {\bf{r}} \}) }
\end{eqnarray}
so that Eq.~\ref{Equ:force_2} becomes 
\begin{eqnarray}
\beta {\bf f}_{{\bf r}_i} &=& \sum_{j=1}^{N_\mathrm{p}}  \left[
 { {\bf \nabla}_{{\bf r}_i} \Omega_{j}(\{ {\bf{r}} \}) \over \Omega_{j}(\{ {\bf{r}} \}) } \left( \on^\mathrm{A}_j + \on^\mathrm{B}_j + \on^{\mathrm{AB}}_{jj} \right)
+ { {\bf \nabla}_{{\bf r}_i} \Omega^\mathrm{s}_{j}(\{ {\bf{r}} \}) \over \Omega^\mathrm{s}_{j}(\{ {\bf{r}} \}) } \on^{\mathrm{AC}}_j
\right]
\nonumber \\
&& +\sum_{1\leq j < q \leq N_\mathrm{p}} { {\bf \nabla}_{{\bf r}_i} \Omega_{jq}(\{ {\bf{r}} \}) \over \Omega_{jq}(\{ {\bf{r}} \}) } \left( \on^{\mathrm{AB}}_{jq} + \on^{\mathrm{BA}}_{jq} \right)
+\on^\mathrm{C}_\mathrm{s} { {\bf \nabla}_{{\bf r}_i} \Omega^\mathrm{s}(\{ {\bf{r}} \}) \over \Omega^\mathrm{s}(\{ {\bf{r}} \}) }
 - \beta {\bf \nabla}_{{\bf r}_i} V(\{ {\bf{r}} \})
\label{Equ:force_3}
\end{eqnarray}
Using Main Eqs.~11, along with the fact that $\Omega^\mthr{s}_j$ and $\Omega_{jq}$ are only function, respectively, of ${\bf r}_j$ and $\{ {\bf r}_j,\, {\bf r}_q \}$ we find 
\begin{eqnarray}
\beta {\bf f}_i &=& \sum_{j \in v(i)} 
\left[
\left(   \on^{\mathrm{AB}}_{ij} + \on^{\mathrm{BA}}_{ij} \right) { {\bf \nabla}_{{\bf r}_i} \Omega_{ij}( r_{ij} ) \over \Omega_{ij}( r_{ij} ) }
-\left( \on^\mathrm{A}_j + \on^\mathrm{B}_j + \on^{\mathrm{AB}}_{jj} \right) {{\bf \nabla}_{{\bf r}_i} e_{ji}(r_{ij}) \over \Omega_j(\{{\bf{r}}\}) }
\right.
\nonumber \\
&& \left.  -\left( \on^\mathrm{A}_i + \on^\mathrm{B}_i + \on^{\mathrm{AB}}_{ii} \right) {{\bf \nabla}_{{\bf r}_i} e_{ij}(r_{ij}) \over \Omega_i(\{ {\bf{r}}\}) }
\right] + \on^{\mathrm{AC}}_i { {\bf \nabla}_{{\bf r}_i} \Omega^\mathrm{s}_{i}( r_{i,z} ) \over w^\mathrm{s}_{i}( r_{i,z}) }
\nonumber \\
&& - \on^\mathrm{C}_\mathrm{s} { {\bf \nabla}_{{\bf r}_i} k^\mathrm{s}_i(r_{i,z}) \over \Omega^\mathrm{s}(\{ {\bf{r}} \}) } 
 -\left( \on^\mathrm{A}_i + \on^\mathrm{B}_i + \on^{\mathrm{AB}}_{ii} \right) {{\bf \nabla}_{{\bf r}_i} e^\mathrm{s}_i(r_{i,z}) \over \Omega_i(\{ {\bf{r}}\}) }
- \beta {\bf \nabla}_i V(\{ {\bf{r}} \})
\label{Equ:force_4}
\end{eqnarray}
When using the EXP algorithm (see Main Sec.~2.2), the linkages are evolved using the Gillespie algorithm (see next section). In this case, ${\bf f}_i$ is calculated as in the last equality of Eq.~\ \ref{Equ:force_1} in view of the fact that $\{ n \}$ are not a function of $\{ {\bf r} \}$. Therefore, the expression of ${\bf f}_i$ used in the EXP case is identical to Eq.~\ref{Equ:force_4} when replacing $\{ \on \}$ with the actual values of the linkages $\{ n \}$.
\\
Concerning the grand-canonical Monte Carlo algorithm, insertion/removal acceptances are given by 
\begin{eqnarray}
\mathrm{acc}_\mthr{ins} &=& \mathrm{min}\left[1, \frac{V}{(N_\mathrm{p}+1)}\cdot\rho_\mathrm{id}\cdot\exp\left[-\beta \Delta{\cal F}_{\mathrm{ins}}\right] \right]
\nonumber \\
\mathrm{acc}_\mthr{rem} &=& \mathrm{min}\left[1, \frac{N_\mathrm{p}}{V}\cdot\frac{1 }{ \rho_\mathrm{id}}\cdot\exp\left[-\beta \Delta{\cal F}_\mathrm{rem}\right] \right]
\end{eqnarray}
where $\Delta{\cal F}_\mathrm{rem}$/$\Delta{\cal F}_\mathrm{ins}$ is the change of the system free energy after removing/inserting a colloid in the simulation box. To optimize the acceptances, we set $\Delta{\cal F}_\mathrm{rem}$ and $\Delta{\cal F}_\mathrm{ins}$ to zero when the inserted/removed particle does not interact with any other particle or the surface.

\subsection*{S3.A: Gillespie algorithm}
In this section, we detail the implementation of the Gillespie algorithm that has been used to simulate sticky-ends reactions in the EXP method.
\\
At a given colloid configuration, $\{ {\bf r} \}$, we start calculating all $on$/$off$ rates of making/breaking linkages
\begin{eqnarray}
 k_\mathrm{on}^{ij}, 
 \qquad
k_\mathrm{on}^{ii},
 \qquad
k_\mathrm{on}^{i,\mthr{s}},
 \qquad
 k_\mathrm{off}^{ij},
 \qquad
 k_\mathrm{off}^{ii},
 \qquad
 k_\mathrm{off}^{i,\mthr{s}}, 
\nonumber
\end{eqnarray}
as derived in Main Eq.~8. Notice that, for instance, $k_\mathrm{on}^{ij}$ is the $on$ rate of making a bridge between $i$ and $j$ either using an A sticky end tethered to particle $i$ or $j$. While $off$ rates are only function of $\Delta G_0$ or $\Delta G^\mthr{s}_0$, $on$ rates also include configurational terms that are function of $\{ {\bf r} \}$ (see Main Eq.~8).
Accordingly, the list of all possible reactions is specified by the following affinities
\begin{eqnarray}
a_\mathrm{on,AB}^{ij}= n^\mthr{A}_i n^\mthr{B}_j k_\mathrm{on}^{ij}, 
 \quad
 a_\mathrm{on,BA}^{ij}= n^\mthr{B}_i n^\mthr{A}_j k_\mathrm{on}^{ij}, 
 \quad
a_\mathrm{on}^{ii}= n^\mthr{A}_i n^\mthr{B}_i k_\mathrm{on}^{ii},
 \quad
a_\mathrm{on}^{i,\mthr{s}}= n^\mthr{A}_i n^\mthr{C}_\mathrm{s} k_\mathrm{on}^{i,\mthr{s}},
 \nonumber \\
a_\mathrm{off,AB}^{ij}=n^\mthr{AB}_{ij} k_\mathrm{off}^{ij},
 \quad
 a_\mathrm{off,BA}^{ij}=n^\mthr{BA}_{ij} k_\mathrm{off}^{ij},
 \quad
 a_\mathrm{off}^{ii}=n_{ii} k_\mathrm{off}^{ii},
 \quad
a_\mathrm{off}^{i,\mthr{s}}=n^\mthr{AC}_i k_\mathrm{off}^{i,\mthr{s}},
\label{Eq:affinities}
\end{eqnarray}
where, for instance, $a_\mathrm{on,AB}^{ij}$ refers to the possibility of forming a linkage between $i$ and $j$ using an A sticky end tethered to particle $i$. 
We then fire one within all possible reactions with probability 
\begin{eqnarray}
p_\mathrm{on,AB}^{ij}= {n^\mthr{A}_i n^\mthr{B}_j k_\mathrm{on}^{ij}\over a_\mthr{tot} }, 
 \quad
 p_\mathrm{on,BA}^{ij}={ n^\mthr{B}_i n^\mthr{A}_j k_\mathrm{on}^{ij}\over a_\mthr{tot} }, 
 \quad
p_\mathrm{on}^{ii}={ n^\mthr{A}_i n^\mthr{B}_i k_\mathrm{on}^{ii}\over a_\mthr{tot} },
 \quad
p_\mathrm{on}^{i,\mthr{s}}={ n^\mthr{A}_i n^\mthr{C}_\mathrm{s} k_\mathrm{on}^{i,\mthr{s}} \over a_\mthr{tot} },
 \nonumber \\
p_\mathrm{off,AB}^{ij}={ n^\mthr{AB}_{ij} k_\mathrm{off}^{ij} \over a_\mthr{tot} },
 \quad
 p_\mathrm{off,BA}^{ij}={ n^\mthr{BA}_{ij} k_\mathrm{off}^{ij} \over a_\mthr{tot} },
 \quad
 p_\mathrm{off}^{ii}={n_{ii} k_\mathrm{off}^{ii} \over a_\mthr{tot} },
 \quad
p_\mathrm{off}^{i,\mthr{s}}={n^\mthr{AC}_i k_\mathrm{off}^{i,\mthr{s}} \over a_\mthr{tot} },
\end{eqnarray}
where $ a_\mthr{tot}$ is the total affinity. Along with the type of reaction we sample the time for it to happen ($\tau$), distributed as $P(\tau) = \exp[-\tau/a_\mthr{tot}]/a_\mthr{tot}$, and increment a reaction clock $\tau_\mthr{reac}$ by $\tau$. If $\tau_\mthr{reac}< \Delta t$ (where $\Delta t$ is the simulation step), we update $\{ n \}$, recalculate the affinities (Eq.~\ref{Eq:affinities}), and fire a new reaction until reaching $\Delta t$.

\end{widetext}

\end{document}